\documentclass[prd,aps,floats,twocolumn,nofootinbib,superscriptaddress]{revtex4-2}
\usepackage{natbib}
\usepackage{commath}
\usepackage{makecell}
\pdfoutput=1
\usepackage[usenames, dvipsnames]{color}
\usepackage[T1]{fontenc}
\usepackage{lmodern}
\usepackage{graphicx, array}
\usepackage{multirow}
\usepackage{graphics}
\usepackage{graphics,epsfig}
\usepackage{amssymb}
\usepackage{amsmath}
\usepackage{ulem}
\usepackage{multirow}
\usepackage{subfigure}
\usepackage{bm}
\usepackage{txfonts}
\usepackage{cancel}
\usepackage{url}
\usepackage{hyperref}
\usepackage{lipsum}
\usepackage{graphicx}
\usepackage{comment} 
\usepackage{soul}

\def\aj{Astronomical Journal}%
\newcolumntype{C}[1]{>{\centering\arraybackslash}m{#1}}

\def\be{\begin{equation}}
\def\ee{\end{equation}}
\def\bi{\begin{itemize}}
\def\ei{\end{itemize}}
\def\ben{\begin{enumerate}}
\def\een{\end{enumerate}}
\def\bt{\begin{tabular}}
\def\et{\end{tabular}}
\def\bc{\begin{center}}
\def\ec{\end{center}}

\def\bea{\begin{eqnarray}}
\def\eea{\end{eqnarray}}

\def\ba{\begin{eqnarray}}
\def\ea{\end{eqnarray}}

\renewcommand{\arraystretch}{1.2}

\let\oldhat\hat

\renewcommand{\hat}[1]{\oldhat{\boldsymbol{\mathbf{#1}}}}

\graphicspath{ {./images/},{ ./images/ratioFigs/},{ ./images/windowFigs/},{ ./Version1/}}

\begin{document}



\title{Probing Baryons with the Kinematic Sunyaev--Zel'dovich Effect and \\ Machine Learning-Derived Peculiar Velocities using DESI DR2 and ACT DR6}

\author{Yulin Gong}
\affiliation{Department of Astronomy, Cornell University, Ithaca, NY 14853, USA}
\author{Rachel Bean}
\affiliation{Department of Astronomy, Cornell University, Ithaca, NY 14853, USA}
\author{Patricio A. Gallardo} 
\affiliation{Department of Physics and Astronomy, University of Pennsylvania, 209 South 33rd Street, Philadelphia, PA, USA 19104}
\author{Boryana Hadzhiyska}
\affiliation{Institute of Astronomy, Madingley Road, Cambridge, CB3 0HA, UK}
\affiliation{Kavli Institute for Cosmology Cambridge, Madingley Road, Cambridge, CB3 0HA, UK}
\author{Yun-Hsin Hsu}
\affiliation{Institute of Astronomy and Astrophysics, Academia Sinica\,(ASIAA),  Taipei 10617, Taiwan}
\author{Jenna Moore}
\author{Eve M. Vavagiakis}
\affiliation{Department of Physics, Duke University, Durham, NC, 27708}
\author{Nicholas Battaglia}
\affiliation{Department of Astronomy, Cornell University, Ithaca, NY 14853, USA}

\author{J.~Aguilar}
\affiliation{Lawrence Berkeley National Laboratory, 1 Cyclotron Road, Berkeley, CA 94720, USA}
\author{S.~Ahlen}
\affiliation{Department of Physics, Boston University, 590 Commonwealth Avenue, Boston, MA 02215 USA}
\author{A.~Aviles}
\affiliation{Instituto Avanzado de Cosmolog\'{\i}a A.~C., San Marcos 11 - Atenas 202. Magdalena Contreras. Ciudad de M\'{e}xico C.~P.~10720, M\'{e}xico}
\affiliation{Instituto de Ciencias F\'{\i}sicas, Universidad Nacional Aut\'onoma de M\'exico, Av. Universidad s/n, Cuernavaca, Morelos, C.~P.~62210, M\'exico}
\author{F.~Beutler}
\affiliation{Institute for Astronomy, University of Edinburgh, Royal Observatory, Blackford Hill, Edinburgh EH9 3HJ, UK}
\author{D.~Bianchi}
\affiliation{Dipartimento di Fisica ``Aldo Pontremoli'', Universit\`a degli Studi di Milano, Via Celoria 16, I-20133 Milano, Italy}
\affiliation{INAF-Osservatorio Astronomico di Brera, Via Brera 28, 20122 Milano, Italy}
\author{D.~Brooks}
\affiliation{Department of Physics \& Astronomy, University College London, Gower Street, London, WC1E 6BT, UK}
\author{A.~Carnero Rosell}
\affiliation{Departamento de Astrof\'{\i}sica, Universidad de La Laguna (ULL), E-38206, La Laguna, Tenerife, Spain}
\affiliation{Instituto de Astrof\'{\i}sica de Canarias, C/ V\'{\i}a L\'{a}ctea, s/n, E-38205 La Laguna, Tenerife, Spain}
\author{T.~Claybaugh}
\affiliation{Lawrence Berkeley National Laboratory, 1 Cyclotron Road, Berkeley, CA 94720, USA}
\author{A.~de la Macorra}
\affiliation{Instituto de F\'{\i}sica, Universidad Nacional Aut\'{o}noma de M\'{e}xico,  Circuito de la Investigaci\'{o}n Cient\'{\i}fica, Ciudad Universitaria, Cd. de M\'{e}xico  C.~P.~04510,  M\'{e}xico}
\author{Arjun~Dey}
\affiliation{NSF NOIRLab, 950 N. Cherry Ave., Tucson, AZ 85719, USA}
\author{Biprateep~Dey}
\affiliation{Department of Astronomy \& Astrophysics, University of Toronto, Toronto, ON M5S 3H4, Canada}
\affiliation{Department of Physics \& Astronomy and Pittsburgh Particle Physics, Astrophysics, and Cosmology Center (PITT PACC), University of Pittsburgh, 3941 O'Hara Street, Pittsburgh, PA 15260, USA}
\author{P.~Doel}
\affiliation{Department of Physics \& Astronomy, University College London, Gower Street, London, WC1E 6BT, UK}
\author{A.~Font-Ribera}
\affiliation{Instituci\'{o} Catalana de Recerca i Estudis Avancats, Passeig de Llu\'{\i}s Companys, 23, 08010 Barcelona, Spain}
\affiliation{Institut de F\'{i}sica d’Altes Energies (IFAE), The Barcelona Institute of Science and Technology, Edifici Cn, Campus UAB, 08193, Bellaterra (Barcelona), Spain}
\author{J.~E.~Forero-Romero}
\affiliation{Departamento de F\'isica, Universidad de los Andes, Cra. 1 No. 18A-10, Edificio Ip, CP 111711, Bogot\'a, Colombia}
\affiliation{Observatorio Astron\'omico, Universidad de los Andes, Cra. 1 No. 18A-10, Edificio H, CP 111711 Bogot\'a, Colombia}
\author{E.~Gaztañaga}
\affiliation{Institut d'Estudis Espacials de Catalunya (IEEC), c/ Esteve Terradas 1, Edifici RDIT, Campus PMT-UPC, 08860 Castelldefels, Spain}
\affiliation{Institute of Cosmology and Gravitation, University of Portsmouth, Dennis Sciama Building, Portsmouth, PO1 3FX, UK}
\affiliation{Institute of Space Sciences, ICE-CSIC, Campus UAB, Carrer de Can Magrans s/n, 08913 Bellaterra, Barcelona, Spain}
\author{G.~Gutierrez}
\affiliation{Fermi National Accelerator Laboratory, PO Box 500, Batavia, IL 60510, USA}
\author{K.~Honscheid}
\affiliation{Center for Cosmology and AstroParticle Physics, The Ohio State University, 191 West Woodruff Avenue, Columbus, OH 43210, USA}
\affiliation{Department of Physics, The Ohio State University, 191 West Woodruff Avenue, Columbus, OH 43210, USA}
\affiliation{The Ohio State University, Columbus, 43210 OH, USA}
\author{S.~Juneau}
\affiliation{NSF NOIRLab, 950 N. Cherry Ave., Tucson, AZ 85719, USA}
\author{T.~Karim}
\affiliation{Department of Astronomy \& Astrophysics, University of Toronto, Toronto, ON M5S 3H4, Canada}
\author{R.~Kehoe}
\affiliation{Department of Physics, Southern Methodist University, 3215 Daniel Avenue, Dallas, TX 75275, USA}
\author{D.~Kirkby}
\affiliation{Department of Physics and Astronomy, University of California, Irvine, 92697, USA}
\author{A.~Kremin}
\affiliation{Lawrence Berkeley National Laboratory, 1 Cyclotron Road, Berkeley, CA 94720, USA}
\author{O.~Lahav}
\affiliation{Department of Physics \& Astronomy, University College London, Gower Street, London, WC1E 6BT, UK}
\author{M.~Landriau}
\affiliation{Lawrence Berkeley National Laboratory, 1 Cyclotron Road, Berkeley, CA 94720, USA}
\author{L.~Le~Guillou}
\affiliation{Sorbonne Universit\'{e}, CNRS/IN2P3, Laboratoire de Physique Nucl\'{e}aire et de Hautes Energies (LPNHE), FR-75005 Paris, France}
\author{M.~Manera}
\affiliation{Departament de F\'{i}sica, Serra H\'{u}nter, Universitat Aut\`{o}noma de Barcelona, 08193 Bellaterra (Barcelona), Spain}
\affiliation{Institut de F\'{i}sica d’Altes Energies (IFAE), The Barcelona Institute of Science and Technology, Edifici Cn, Campus UAB, 08193, Bellaterra (Barcelona), Spain}
\author{A.~Meisner}
\affiliation{NSF NOIRLab, 950 N. Cherry Ave., Tucson, AZ 85719, USA}
\author{R.~Miquel}
\affiliation{Instituci\'{o} Catalana de Recerca i Estudis Avancats, Passeig de Llu\'{\i}s Companys, 23, 08010 Barcelona, Spain}
\affiliation{Institut de F\'{i}sica d’Altes Energies (IFAE), The Barcelona Institute of Science and Technology, Edifici Cn, Campus UAB, 08193, Bellaterra (Barcelona), Spain}
\author{S.~Nadathur}
\affiliation{Institute of Cosmology and Gravitation, University of Portsmouth, Dennis Sciama Building, Portsmouth, PO1 3FX, UK}
\author{J.~ A.~Newman}
\affiliation{Department of Physics \& Astronomy and Pittsburgh Particle Physics, Astrophysics, and Cosmology Center (PITT PACC), University of Pittsburgh, 3941 O'Hara Street, Pittsburgh, PA 15260, USA}
\author{H.~E.~Noriega}
\affiliation{Instituto de Ciencias F\'{\i}sicas, Universidad Nacional Aut\'onoma de M\'exico, Av. Universidad s/n, Cuernavaca, Morelos, C.~P.~62210, M\'exico}
\affiliation{Instituto de F\'{\i}sica, Universidad Nacional Aut\'{o}noma de M\'{e}xico,  Circuito de la Investigaci\'{o}n Cient\'{\i}fica, Ciudad Universitaria, Cd. de M\'{e}xico  C.~P.~04510,  M\'{e}xico}
\author{E.~Paillas}
\affiliation{Instituto de Estudios Astrof\'isicos, Facultad de Ingenier\'ia y Ciencias, Universidad Diego Portales, Av. Ej\'ercito Libertador 441, Santiago, Chile}
\affiliation{Steward Observatory, University of Arizona, 933 N. Cherry Avenue, Tucson, AZ 85721, USA}
\author{N.~Palanque-Delabrouille}
\affiliation{IRFU, CEA, Universit\'{e} Paris-Saclay, F-91191 Gif-sur-Yvette, France}
\affiliation{Lawrence Berkeley National Laboratory, 1 Cyclotron Road, Berkeley, CA 94720, USA}
\author{W.~J.~Percival}
\affiliation{Department of Physics and Astronomy, University of Waterloo, 200 University Ave W, Waterloo, ON N2L 3G1, Canada}
\affiliation{Perimeter Institute for Theoretical Physics, 31 Caroline St. North, Waterloo, ON N2L 2Y5, Canada}
\affiliation{Waterloo Centre for Astrophysics, University of Waterloo, 200 University Ave W, Waterloo, ON N2L 3G1, Canada}
\author{F.~Prada}
\affiliation{Instituto de Astrof\'{i}sica de Andaluc\'{i}a (CSIC), Glorieta de la Astronom\'{i}a, s/n, E-18008 Granada, Spain}
\author{I.~P\'erez-R\`afols}
\affiliation{Departament de F\'isica, EEBE, Universitat Polit\`ecnica de Catalunya, c/Eduard Maristany 10, 08930 Barcelona, Spain}
\author{C.~Ravoux}
\affiliation{Universit\'{e} Clermont-Auvergne, CNRS, LPCA, 63000 Clermont-Ferrand, France}
\author{G.~Rossi}
\affiliation{Department of Physics and Astronomy, Sejong University, 209 Neungdong-ro, Gwangjin-gu, Seoul 05006, Republic of Korea}
\author{L.~Samushia}
\affiliation{Abastumani Astrophysical Observatory, Tbilisi, GE-0179, Georgia}
\affiliation{Department of Physics, Kansas State University, 116 Cardwell Hall, Manhattan, KS 66506, USA}
\author{E.~Sanchez}
\affiliation{CIEMAT, Avenida Complutense 40, E-28040 Madrid, Spain}
\author{C.~Saulder}
\affiliation{Max Planck Institute for Extraterrestrial Physics, Gie\ss enbachstra\ss e 1, 85748 Garching, Germany}
\author{D.~Schlegel}
\affiliation{Lawrence Berkeley National Laboratory, 1 Cyclotron Road, Berkeley, CA 94720, USA}
\author{M.~Schubnell}
\affiliation{Department of Physics, University of Michigan, 450 Church Street, Ann Arbor, MI 48109, USA}
\affiliation{University of Michigan, 500 S. State Street, Ann Arbor, MI 48109, USA}
\author{H.~Seo}
\affiliation{Department of Physics \& Astronomy, Ohio University, 139 University Terrace, Athens, OH 45701, USA}
\author{J.~Silber}
\affiliation{Lawrence Berkeley National Laboratory, 1 Cyclotron Road, Berkeley, CA 94720, USA}
\author{M.~Siudek}
\affiliation{Institute of Space Sciences, ICE-CSIC, Campus UAB, Carrer de Can Magrans s/n, 08913 Bellaterra, Barcelona, Spain}
\affiliation{Instituto de Astrof\'{\i}sica de Canarias, C/ V\'{\i}a L\'{a}ctea, s/n, E-38205 La Laguna, Tenerife, Spain}
\author{G.~Tarl\'{e}}
\affiliation{University of Michigan, 500 S. State Street, Ann Arbor, MI 48109, USA}
\author{B.~A.~Weaver}
\affiliation{NSF NOIRLab, 950 N. Cherry Ave., Tucson, AZ 85719, USA}
\author{R.~Zhou}
\affiliation{Lawrence Berkeley National Laboratory, 1 Cyclotron Road, Berkeley, CA 94720, USA}

\date{\today}

\begin{abstract}
We present novel empirical constraints on average optical depths and gas density profiles from combining kinematic Sunyaev–Zel’dovich effect (kSZ) measurements and nonlinear peculiar 
velocities, reconstructed using machine learning, for massive halos traced by DESI DR2 luminous red galaxies (LRG).
The pairwise kSZ is measured using two ACT DR6 cosmic microwave background (CMB) temperature maps (ILC and 150 GHz maps) for seven luminosity-selected LRG samples. We find consistent results between the two maps and significant detections across the samples. We reconstruct LRG line-of-sight peculiar velocities in two cosmic epochs, centered on $z=$ 0.55 and 0.8, using a simulation-trained Transformer machine learning model to extend beyond the linear approximation.  Average optical depths are inferred by comparing the kSZ pairwise correlation to the measured pairwise velocity correlation from the reconstructed velocity field and, separately, from a theoretical prediction using the best fit Planck cosmology, with the highest significance reaching SNR = 14.2 and 15.1 respectively.  We also obtain velocity-weighted AP-filtered kSZ-stacked cluster density profiles, which show evidence of extended ionized gas around massive LRG groups and no strong redshift evolution within current measurement uncertainties. These results demonstrate the power of combining spectroscopic galaxy and CMB data, and employing machine learning methods, to probe diffuse baryons and coherent large-scale velocity fields. They open up new avenues to leverage upcoming galaxy survey data from Euclid, Roman and Rubin LSST, in tandem with multi-frequency CMB/sub-mm data from CCAT and Simons Observatory. 
\end{abstract}

\maketitle
 
\section{Introduction}
\label{sec:intro}
Understanding the distribution of baryons around galaxy groups and clusters is a central problem in modern cosmology \citep{Fukugita:2003gm,Tumlinson_2017}. While the total abundance of baryons in the Universe is tightly constrained by cosmic microwave background (CMB) and Big Bang nucleosynthesis measurements \citep{Planck:2018vyg,AtacamaCosmologyTelescope:2025blo,AtacamaCosmologyTelescope:2025nti}, a large fraction of baryons is expected to reside in diffuse ionized gas outside galaxies \citep{Cen:2006by,Shull_2012}. This gas is shaped by gravitational collapse, accretion, star formation, and feedback from stars and active galactic nuclei, and therefore encodes key information about both galaxy structure formation and evolution \citep{Nelson:2014aea,Suresh_2015}. An inaccurate description of baryonic physics may systematically bias constraints on important cosmological parameters, such as $\sigma_8$ and the sum of neutrino mass \citep{Harnois-Deraps:2014sva,Parimbelli:2018yzv,Schneider:2019snl,Amon:2022azi, DESI:2024hhd,DESI:2025zgx}.

The interaction of CMB photons with hot cluster gas imprints two secondary Sunyaev-Zel'dovich (SZ) signatures onto the primary CMB temperature fluctuations  \cite{Sunyaev:1972eq,Sunyaev:1980nv}. The thermal SZ (tSZ) effect, from inverse Compton scattering of the CMB off the free electrons in the hot gas, induces a frequency-dependent modification that can be isolated using multi-frequency CMB measurements. The kinematic Sunyaev-Zel'dovich effect (kSZ) arises from the Doppler shifting of the CMB due to the bulk peculiar velocities of the gas. It is smaller in amplitude than the tSZ, and largely frequency-independent, but provides a powerful and complementary probe of the density and velocity of the diffuse ionized scattering gas. Unlike tSZ and X-ray profiles which are dominated by hot central regions of the cluster, the kSZ signal is insensitive to the gas temperature and is therefore well suited to probe the distribution of lower temperature baryons in the outskirts of the cluster \citep{ACTPol:2015teu,Ferraro:2016ymw}.
 
Beyond its role as a probe of diffuse baryons, the kSZ effect also has important cosmological implications because it directly traces the peculiar velocity field sourced by the growth of large-scale structure. kSZ-based velocity measurements therefore provide an independent way to test the standard cosmological model and modified gravity \cite{Clifton:2011jh,Joyce:2014kja,Joyce:2016vqv,Nojiri:2017ncd,ACT:2026rfx}. Since peculiar velocities are sensitive to both the amplitude and the rate of structure growth, pairwise statistics can also provide complementary constraints on massive neutrinos, the growth rate of structure, and dark energy \cite{DeDeo:2005yr,Bhattacharya:2007sk,Kosowsky_2009,Mueller:2014dba,Mueller:2014nsa}.

Over the past decade, kSZ measurements have developed from first detections into a rapidly growing probe of both baryonic gas and cosmic velocities \cite{Hand:2012ui,Smith:2018bpn,Vavagiakis:2021ilq,McCarthy:2025brx,EmbilVillagra:2026fnh,Qu:2026zyh,Hadzhiyska:2026uyq,Chaussidon:2026vmn}. Among the different measurements, pairwise kSZ statistics use the mean gravitational infall of cluster pairs to detect the average relative motion of ionized gas. Since, on large scales, galaxy clusters are expected to move toward each other, their kSZ temperature differences contain a characteristic separation-dependent signal.

Different galaxy samples can probe different physical regimes. Luminous red galaxies (LRGs) are tracers with high bias that typically occupy higher mass halos, associated with galaxy groups and clusters, \citep{Zheng:2008np,SDSS:2006egz}, making them especially suitable for pairwise kSZ measurements \cite{DeBernardis:2016pdv}. Using spectroscopically-selected LRGs as tracers of the centers of massive halos, the pairwise kSZ approach has provided a direct probe for the large scale motion of ionized gas \cite{DES:2016umt,Calafut:2021wkx,Hadzhiyska:2025egz,Gong:2025ffw}. In addition to pairwise estimators, a complementary class of measurements uses reconstructed peculiar velocities to stack CMB temperature maps around galaxy clusters \cite{AtacamaCosmologyTelescope:2020wtv,DES:2023mug,Hadzhiyska:2024qsl,Qu:2026zyh}, enabling measurements of the projected electron density and optical depth profiles. These velocity-weighted kSZ profiles provide a direct probe of the gas distribution around galaxy clusters and can be compared across different galaxy samples, redshift ranges, and halo masses. Projected field methods have also been actively studied as a complementary way to extract the kSZ signal through correlation with a variety of large scale structure tracers, including emission line galaxies, and without requiring individual spectroscopic redshifts \cite{Hill:2016dta,Patki:2023fjz,Rodriguez:2025jll,Patki:2025oyp}.

The combination of large sky area spectroscopic surveys and high resolution CMB maps has made these measurements increasingly powerful. The Dark Energy Spectroscopic Instrument (DESI) \cite{DES:2016umt} provides precise redshifts for millions of galaxies and quasars over a large fraction of the sky, enabling dense three-dimensional maps of large scale structure. In parallel, the Atacama Cosmology Telescope (ACT) \cite{Swetz_2011} provides high-resolution CMB temperature maps with broad overlap with the DESI survey. The overlap between DESI galaxies and ACT DR6 CMB data \cite{AtacamaCosmologyTelescope:2025vnj} is particularly well suited for kSZ studies. DESI supplies the redshift information and large scale structure tracers needed to estimate velocities and construct pairwise statistics, while ACT provides the CMB temperature maps from which the kSZ signal can be measured.

A key challenge in probing baryons with kSZ is that the observed signal depends on both the electron optical depth and the peculiar velocity. As a result, constraints on the ionized gas content of halos require either an external velocity model or an independent velocity estimate. One way is to compare the measured pairwise kSZ signal to a theoretical pairwise velocity prediction from linear theory to constrain an effective optical depth (e.g. \cite{Calafut:2021wkx}). A second approach can use reconstructed peculiar velocities to calculate the pairwise velocity and compare it with the kSZ (e.g. \cite{Ma:2017ybt}). A third complementary approach can use stacked tSZ measurements to constrain optical depth around the same galaxy samples and provide an independent comparison for kSZ-derived optical depth constraints (e.g. \cite{Vavagiakis:2021ilq,Moore:2026inprep}).

A key ingredient for both pairwise velocity and velocity-weighted kSZ profile measurements is the reconstruction of the line of sight (LOS) peculiar velocity. Standard approaches often rely on the analytic linearized continuity equation to infer velocities from the observed galaxy density field \citep{Guachalla:2023lbx,Hadzhiyska:2023nig}. While these methods are physically motivated and have been widely used, they are primarily sensitive to large-scale linear modes and therefore miss part of the nonlinear velocity information that becomes important for accurate kSZ calibration and profile measurements. 

Machine learning methods have become increasingly common in astronomy and cosmology to move beyond analytic solutions to model complex facets of the data. For example, they have used to estimate photometric redshifts, classify galaxy morphologies, identify gravitational lenses, and analyze large-scale sky-survey data \cite{baron2019machinelearningastronomypractical,2022,Huertas-Company:2022wni}. Recent reviews also emphasize that machine learning has become broadly used across astrophysical data analysis, especially because modern surveys produce increasingly large and complex datasets \cite{Soo:2023avq}. Machine learning methods have recently been applied directly to the reconstruction of cosmological density and velocity fields \cite{Veena:2022now,Lilow:2024rwa,Xiao:2024gcz} to improve the cosmological analysis. Motivated by these developments, we use the machine learning to reconstruct peculiar velocity in this work.

We use the physically-motivated linear continuity equation reconstruction as a baseline and build on it using machine learning to predict the nonlinear residual between the true LOS peculiar velocity and the linearly reconstructed velocity \citep{Gong:2026xld}. This residual learning strategy preserves the robustness and interpretability of the standard reconstruction on large scales, while allowing the model to recover nonlinear corrections that are not captured by the linear continuity equation. In \citep{Gong:2026xld}, both gradient boosting decision trees and Transformer-based models were trained on AbacusSummit mock DESI LRG and ELG catalogs to reconstruct nonlinear velocity residuals using multi-scale displacement and density-environment features. The Transformer model was shown to provide the strongest performance in tests with simulated data: improving the correlation with the true velocity field, recovering the velocity power spectrum more accurately over a wider range of scales, and enhancing downstream kSZ-related applications such as pairwise velocity statistics and velocity-weighted stacked density profiles. In this paper, we apply the Transformer model to analyze the real DESI LRG data with the goal of reconstructing accurate LOS velocities and combining them with ACT kSZ CMB temperature measurements to gain a deeper insight into the baryons within galaxy clusters.

In this paper, we measure the pairwise kSZ momentum, the galaxy cluster pairwise velocity statistic, and velocity-weighted cluster gas density profiles using DESI Data Release 2 (DR2) galaxies and ACT Data Release 6 (DR6) CMB data. Together, these observables probe both the baryonic gas associated with massive halos and the coherent velocity field traced by LRGs. This analysis builds directly on the previous ACT DR6 × DESI DR1 kSZ measurement presented in \cite{Gong:2025ffw}, which reported a high-significance detection of the pairwise kSZ signal using DESI DR1 LRGs together with ACT DR6 and Planck CMB temperature maps. That work established the effectiveness of combining DESI spectroscopic galaxies with high-resolution CMB data to measure the coherent pairwise motion of ionized gas. Here we extend that framework to the larger DESI DR2 LRG sample and use the increased statistical power to improve the pairwise kSZ measurement, test the luminosity and halo-mass dependence of the AP-filtered effective optical depth. We also combine the improved kSZ measurements with machine learning reconstructed nonlinear peculiar velocities for the LRGs. This allows us to provide the first direct comparisons of pairwise kSZ and pairwise velocity measurements to extract out optical depth estimates and obtain nonlinear velocity-weighted stacked kSZ profiles in two redshift epochs. In this sense, the present work provides both an improvement of the DR1 kSZ pairwise analysis and an explicitly new direction, applying kSZ measurements from increasingly large spectroscopic galaxy samples to characterize diffuse cosmological baryons.

The paper is organized as follows. Section~\ref{sec:data} describes the DESI DR2 galaxy sample, ACT DR6 CMB maps, and masks used in this work. Section~\ref{sec:method} summarizes the  analysis formalism. The results are presented in Section~\ref{sec:results}, including: the  pairwise kSZ statistics, pairwise velocity reconstruction, optical depth estimation and velocity-stacked density profiles. We summarize our conclusions and the implications for future research in Section~\ref{sec:conclusion}.

\section{Data}
\label{sec:data}

\subsection{ACT DR6 CMB maps}
We use CMB temperature maps from the Atacama Cosmology Telescope Data Release 6 (ACT DR6) \citep{AtacamaCosmologyTelescope:2025vnj}. ACT observed the millimeter sky at arcminute resolution from Cerro Toco in northern Chile \cite{Thornton:2016wjq}, with DR6 combining data from multiple frequency bands and observing seasons, from 2017 to 2022. The ACT maps used  overlap substantially with the DESI footprint, providing a well-suited CMB data set for kSZ measurements. The ACT maps are combined with Planck data \citep{Lamarre:2003zh}, spanning 19,000 square degrees of the sky.

Our analysis uses two ACT DR6 temperature products. The first is the {\it f150} frequency map \cite{ACT:2020frw}, centered at 150 GHz, where the CMB signal is large and the tSZ spectral contribution is relatively small compared to lower frequency ones. The second is the internal linear combination map (ILC) \citep{ACT:2024jdf}, a component-separated CMB map that is constructed by combining ACT and external multi-frequency data to suppress foreground emission while preserving the blackbody CMB signal. The ILC and f150 maps contain highly correlated primary CMB and kSZ signals, so we do not expect the ILC map to provide markedly more statistical information than the f150 map alone. We use the ILC map as the fiducial temperature map for our stacked kSZ profile measurements, while the $f150$ map is used as consistency checks for the pairwise kSZ measurements. 

\newcommand{\Lone}{3.6}
\newcommand{\Ltwo}{4.8}
\newcommand{\Lthree}{6.0}
\newcommand{\Lfour}{7.9}
\newcommand{\Lfive}{9.8}

\begin{table}
\centering
\setlength{\tabcolsep}{4pt}
\renewcommand{\arraystretch}{1.25}
\begin{tabular}{|c|c|c|c|c|}
\hline
\multirow{2}{*}{Bin} & Luminosity cut, $L$ & Mass cut, $M$ 
 & \multirow{2}{*}{$N_{\mathrm{gal}}$} & \multirow{2}{*}{$\langle z\rangle$}
\\
 & ($10^{10}L_{\odot}$) & ($10^{13}M_{\odot}$)
&  &
\\ \hline
L36  & $L>\Lone$          & $M>0.38$        &  1,922,899  & 0.76 \\
L48  & $L>\Ltwo$          & $M>0.63$        &  1,453,311  & 0.78 \\
L60  & $L>\Lthree$        & $M>0.96$        & 971,897 & 0.80 \\
L79  & $L>\Lfour$         & $M>1.65$        &  484,090 & 0.84 \\ \hline
L36D & $\Lone<L<\Ltwo$    & $0.38<M<0.63$   &  469,588  & 0.71 \\
L48D & $\Ltwo<L<\Lthree$  & $0.63<M<0.96$   &  481,414  & 0.72 \\
L60D & $\Lthree<L<\Lfour$ & $0.96<M<1.65$   &  487,807  & 0.76 \\
\hline
\end{tabular}
\caption{Summary of the DESI DR2 LRG samples, four cumulative and three disjoint, used in this work. For each sample, we list the bin label, the luminosity cut, the corresponding estimated host halo virial mass proxy, the number of galaxies, $N_{\mathrm{gal}}$, and the average redshift, $\langle z\rangle$. }
\label{tab1}
\end{table}

\subsection{DESI DR2 galaxy catalogs}
We use spectroscopic measurements \cite{DESI:2022nlo} of luminous red galaxies (LRGs) from the DESI DR2 \citep{DESI:2025zpo} catalog. DESI is a multiplexed spectroscopic survey on the Mayall 4-meter telescope at Kitt Peak National Observatory, designed to map the three dimensional distribution of galaxies and quasars over a large fraction of the sky \citep{DESI:2016igz,DESI:2022xcl,DESI:2023iob,DESI:2023mkx,2024AJ....168..245P}. The precise DESI redshifts allow us to reconstruct LOS peculiar velocities for the velocity-weighted stacking analysis and to compare the pairwise kSZ with pairwise velocity measurements.

We select the LRG samples \cite{Zhou:2020mgr,DESI:2022gle,Zhou:2023gji} following the procedure used in the DESI Key Project \citep{DESI:2024uvr} and in the previous ACT DR6 x DESI DR1 \cite{DESI:2025fxa} kSZ analysis \cite{Gong:2025ffw}. We construct the sample from the DESI DR2 spectroscopic product, the LOA redshift catalog. To avoid duplicate objects, we keep only the best available spectrum for each object. We also remove objects with problematic redshift measurements and keep only objects with high confidence redshift fits where we select objects with parameter ZWARN = 0. We further select objects from the main DESI survey observed in the dark time, and identify LRGs using the DESI target selection bitmask. We also apply imaging-quality cuts to remove objects located in regions affected by known imaging problems. We also match the sample to the DESI value added catalog \citep{2023ascl.soft08005M} containing galaxy luminosities and other physical properties. The LRG sample we selected covers a redshift range $0 < z < 1.5$ with peak at $z$ = 0.8.

Before extracting temperatures at galaxy positions, we apply the same mask as used in \cite{Gong:2025ffw}. Specifically, this removes unobserved regions of the ACT DR6 maps. Regions affected by the Galactic plane, bright point sources, and massive clusters whose thermal SZ signal could otherwise dominate the small scale temperature field are also removed. We additionally exclude galaxies located near apodized map boundaries where the map noise is high. These cuts ensure that the galaxy samples used in the kSZ estimators occupy a clean and well-characterized region of the ACT footprint. The effective overlap between the DESI footprint and the ACT DR6 coverage is approximately $7300\,{\rm deg}^2$.

\begin{figure}[t!]
    \centering
    \includegraphics[width=\linewidth]{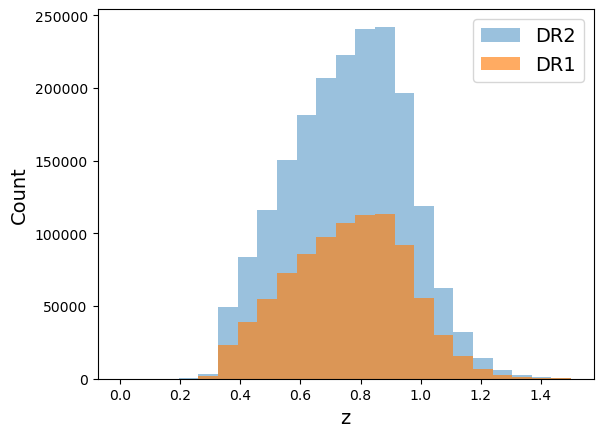}
    \caption{The redshift distributions of the full DESI DR2 L36 sample of 1,922,899 LRGs analyzed in this work [blue] and the  913,286 LRGs used in the previous DESI DR1 analysis \cite{Gong:2025ffw} [orange].
    }
    \label{fig:redshift_dist}
\end{figure}

We construct four cumulative luminosity bins and three disjoint luminosity bins as used in \cite{Gong:2025ffw}. The cumulative samples are labeled L36, L48, L60, and L79, corresponding to luminosity cuts of $L>3.6$, $4.8$, $6.0$, and $7.9$ $\times 10^{10}L_{\odot}$, respectively. The disjoint samples, labeled L36D, L48D, and L60D are formed from the intervals between adjacent luminosity cuts. The final LRG sample after the ACT footprint and mask cuts are summarized in Table~\ref{tab1}. The halo mass is estimated from the stellar masses using the relation in \citep{Kravtsov:2014sra} where the stellar masses are estimated from the luminosity using the relation in \cite{Bell:2003cj}. These estimated halo mass are provided only as approximate physical interpretations of the luminosity-selected samples and are not used as selection criteria to the analysis. We show the redshift distributions of the DR2 LRG sample in Fig.~\ref{fig:redshift_dist}, along with the sample studied in the previous DESI DR1 work \cite{Gong:2025ffw} for comparison. 

We also use the AbacusSummit halo lightcone catalogs \cite{Hadzhiyska:2021uhm} to estimate the pairwise velocity prediction. These catalogs are constructed from the AbacusSummit N-body simulations \cite{Maksimova:2021ynf} which have resolution with $6912^3$ particles in a volume of side length $2\,h^{-1}{Gpc}$. We use the independent 25 realizations, which are based on the Planck 2018 $\Lambda$CDM cosmology \cite{Planck:2018vyg}. The fiducial cosmological parameters are $\omega_b=0.02237$, $\omega_{cdm}=0.1200$, $h=0.6736$, $n_s=0.9649$, and $A_s=2.0830\times10^{-9}$. We create mock LRG catalogs using the AbacusHOD model \cite{Yuan:2023ezi}, a halo occupation distribution (HOD) framework that specifies how galaxies occupy dark matter halos, with model parameters from \cite{Hadzhiyska:2023fic}, calibrated for the DESI survey.

\section{Method}
\label{sec:method}
\subsection{Kinematic Sunyaev--Zel'dovich effect}
\label{sec:ksz}
The kinematic Sunyaev--Zel'dovich (kSZ) effect occurs when CMB photons Thomson scatter off free electrons with nonzero peculiar velocity along the line of sight. The fractional CMB temperature perturbation in the direction $\hat{n}$ is given by
\begin{equation}
    \frac{\Delta T_{kSZ}(\hat{n})}{T_{CMB}}
    =
    - \frac{\sigma_{T}}{c}
    \int dl \,
    n_{e}(r)\,
    v(r) \cdot \hat{n},
    \label{eq:ksz_general}
\end{equation}
where $T_{CMB}$ is the mean CMB temperature, $\sigma_{T}$ is the Thomson cross section, $c$ is the speed of light, $n_{e}$ is the electron number density, and $v\cdot\hat{n}$ is the LOS component of the peculiar velocity.

To measure the kSZ signal from the CMB map, we use aperture photometry (AP) at the cluster positions. The method calculates the average temperature within a circular aperture centered on each cluster, and then subtracts the average temperature in a surrounding annulus with equal area. The AP filtered temperature is defined as
\begin{equation}
    T_{AP}(\theta_{AP})
    =
    \overline{T}_{\theta < \theta_{AP}}
    -
    \overline{T}_{\theta_{AP} < \theta < \sqrt{2}\theta_{AP}},
\end{equation}
where $\theta_{AP}$ is the aperture radius. The filter is therefore defined as the mean temperature within a circular disk minus that in a surrounding annulus with equal area. In this way, the AP filter measures the temperature contrast between the cluster region and its local background. Following \cite{Gong:2025ffw}, we use 2.1$^\prime$ as our aperture size for the analysis in this work. We also subtract a redshift dependent temperature from the measured AP temperature. This correction is intended to remove residual systematic contaminants that could mirror a pairwise signal. This correction term can be written as,
\begin{equation}
    \bar{T}_{AP}(z_i)
    =
    \frac{
    \sum_j T_{AP}
    \exp\left[-(z_i-z_j)^2/(2\sigma_z^2)\right]
    }{
    \sum_j
    \exp\left[-(z_i-z_j)^2/(2\sigma_z^2)\right]
    },
\end{equation}
where $\sigma_z$ = 0.01\cite{Planck:2015ywj}. We have verified using simulation that this redshift dependent subtraction removes temperature contaminants without biasing the pairwise kSZ signal.

\subsection{Pairwise kSZ and velocity statistics}
\label{sec:pairwise}
The pairwise kSZ estimator can be written as
\begin{equation}
    \hat{P}_{kSZ}(r)
    =
    - \frac{
    \sum_{i<j}
    \left(T_{AP,i} - T_{AP,j}\right)c_{ij}
    }{
    \sum_{i<j} c_{ij}^2
    }.
\end{equation}
where the sum is taken over all pairs with comoving separation in the bin $r$. The geometrical weight $c_{ij}$ projects the pair separation along the two lines of sight and accounts for the fact that the observed kSZ signal only contains LOS velocities. It is commonly written as
\begin{eqnarray}
    c_{ij} & \equiv& \hat{r}_{ij}\cdot
\frac{\hat{n}_{i}+\hat{n}_{j}}{2} =\frac{(r_i-r_j)(1+\cos\alpha)}
{2\sqrt{r_i^2+r_j^2-2r_ir_j\cos\alpha}},
\label{eq:cij}
\end{eqnarray}
where $\hat{n}_i$ and $\hat{n}_j$ are the LOS directions to the two clusters, $\hat{r}_{ij}$ is the unit vector between the separation and $\alpha$ is their angular separation.

We measure the pairwise signal in uniform separation bins of width $10$ Mpc, with bin centers from $5$ to $145$ Mpc. We additionally use four wider bins centered at $175$, $225$, $282.5$, and $355$ Mpc. In the following analysis, we mainly focus on scales above $20$ Mpc, where nonlinear velocity contributions are expected to be less significant \cite{Calafut:2021wkx}. We use the package \textsc{Iskay2} developed in \cite{Gallardo_2025} to calculate the pairwise estimator.

Analogous to the kSZ measurement, a pairwise velocity statistic measures the mean relative motion of galaxy or cluster pairs as a function of their comoving separation,
\begin{equation}
    \hat{V}(r)
    =
    -\frac{
    \sum_{i<j}
    \left(v_{\parallel,i} - v_{\parallel,j}\right)c_{ij}
    }{
    \sum_{i<j} c_{ij}^{2}
    },
    \label{eq:Vhat}
\end{equation}
with $v_{\parallel,i}$ is the LOS peculiar velocity of object $i$.

The measured pairwise statistics can be compared with the theoretical prediction from the linear theory \cite{Sheth:2000ff}. For a tracer sample with linear bias $b$, the theoretical pairwise velocity can be written as
\begin{equation} \label{eq:Vhat_theo}
    \hat{V}_{th}(r,z)
    =
    -\frac{2}{3} a(z) H(z) f(z) r
    \frac{b\,\bar{\xi}_{m}(r,z)}
    {1+b^2\xi_m(r,z)} ,
\end{equation}
where $a(z)$ is the scale factor, $H(z)$ is the Hubble parameter, and $f(z)$ is the linear growth rate. The quantity $\xi_m(r,z)$ is the matter two-point correlation function, and $\bar{\xi}_m(r,z)$ is its average value within the volume of radius $r$. For each sample, the linear bias $b$ is estimated following \cite{DES:2016umt}. For the L36 sample, we obtain $b$=2.2, consistent with previous measurements of the DESI LRG bias \cite{Hadzhiyska:2023fic}. The uncertainty in the tracer bias is not propagated through the subsequent analysis and should therefore be regarded as a systematic uncertainty.

In practice, the matter correlation function is computed from the matter power spectrum of the assumed cosmology,
\begin{equation}
    \xi_m(r,z)
    =
    \frac{1}{2\pi^2}
    \int_0^\infty
    k^2 P_m(k,z)
    \frac{\sin(kr)}{kr}
    \,{d}k ,
\end{equation}
where $P_m(k,z)$ is the matter power spectrum at redshift $z$. In this work, we use the best-fit \textit{Planck} cosmological
parameters in \cite{Planck:2018vyg}. The corresponding volume-averaged correlation function is
\begin{equation}
    \bar{\xi}_m(r,z)
    =
    \frac{3}{r^3}
    \int_0^r
    \xi_m(r',z) r'^2\,{d}r' .
\end{equation}

For the kSZ measurement, the theoretical pairwise velocity provides the velocity template that determines the scale dependence of the pairwise kSZ signal. The overall amplitude is set by the effective optical depth of the tracer sample. Therefore, the theoretical pairwise kSZ profile can be modeled as
\begin{equation}
    \hat{P}_{th}(r, \tau)
    =
    \frac{T_{CMB}}{c}\,
    \bar{\tau}\,
    \hat{V}_{th}(r),
    \label{eq:P_theory}
\end{equation}
where $\bar{\tau}$ is the mean effective optical depth. In this work, we compare the measured pairwise kSZ profile with this theoretical template to constrain the mean optical depth of the sample.

\subsection{Velocity reconstruction}
\label{sec:velocity}
A key input for the weighted kSZ profile and the observed pairwise velocity statistic is the LOS peculiar velocity of each tracer. We reconstruct the velocity from the observed DESI galaxy density field. In this approach, the galaxy density contrast is used to infer the large scale displacement field, which is then converted into a peculiar velocity estimate.

We first construct the galaxy overdensity field on a three dimensional mesh and smooth it with a Gaussian kernel. The smoothed density field is then used to infer the displacement field. In Fourier space, the displacement field can be reconstructed as
\begin{equation}
    \Psi(k)
    =
    \frac{ik}{k^2}
    \frac{W_G(k)}
    {b+f\mu^2}
    \delta_g^s(k),
\end{equation}
where $\delta_g^s$ is the observed galaxy overdensity, $b$ is the bias, $f$ is the linear growth rate, $\mu$ is the cosine between the Fourier mode and the line of sight, and $W_G(k)$ is a Gaussian smoothing kernel. We adopt a Gaussian smoothing scale of $R_s = 12.5 h^{-1}$MPC, consistent with that used in \cite{Hadzhiyska:2023nig}. We numerically reconstruct the displacement field using the MultiGrid \cite{White:2015eaa} available in the package \textsc{pyrecon}\footnote{\url{https://github.com/cosmodesi/pyrecon}} \cite{Chen:2024eri}.

The corresponding reconstructed linear velocity is then written as
\begin{equation}
    v_{lin}
    =
    aHf\,\Psi.
\end{equation}

The reconstruction above only captures the large-scale linear velocity field, but it omits the information in nonlinear regime. To improve the velocity estimate, we use the machine learning method developed in \cite{Gong:2026xld}. The model uses the linear reconstruction as a baseline and learns a residual correction between the true and linear reconstructed velocity, $\Delta v_{\mathrm{pred}}$, using multi-scale reconstruction features, including the displacement field and density derived features to predict the missing nonlinear component of the velocity. The reconstructed velocity, $v_{rec}$ is then given by,
\begin{equation}
v_{\mathrm{rec}}
=
v_{\mathrm{lin}}
+
\Delta v_{\mathrm{pred}}.
\label{eq:vfinal}
\end{equation}

Further details of the velocity reconstruction method, including the machine learning architecture, training procedure, and validation on simulations, can be found in \cite{Gong:2026xld}.

Since the kSZ temperature arises from the peculiar motion projected along the line of sight, in this analysis we care about the LOS component $v_{rec\parallel} = v_{rec}\cdot\hat{n}$.

Following the redshift ranges over which the velocity reconstruction model was trained and validated, and using the DESI random clustering catalog available for this analysis, we construct two LRG subsamples from the L36 sample. The first sample covers $0.40 < z < 0.65$ and contains 482,087 galaxies, while the second sample covers $0.65 < z < 0.95$ and contains 971,889 galaxies. The reconstructed velocities allow us to compare a reconstructed pairwise velocity against predictions from theory, in Eq.~(\ref{eq:Vhat_theo}), and from the AbacusSummit simulations. They are also used to estimate the optical depth in Sec.~\ref{sec:tau} and the velocity-weighted kSZ stacking measurement in Sec.~\ref{sec:velocity_profile}. 

\subsection{Mean effective optical depth}
\label{sec:tau}
We estimate $\bar{\tau}_{AP}$ by comparing  the observed pairwise kSZ momentum derived from the AP-filtered kSZ temperature with a prediction modeled based on the optical depth. The best fit optical depth is then obtained by minimizing
\begin{equation}
    \chi^2(\bar{\tau}_{AP})
    =
    \sum_{i,j}
    \Delta \hat{P}_i(\bar{\tau}_{AP})
    C^{-1}_{ij}
    \Delta \hat{P}_j(\bar{\tau}_{AP}) ,
    \label{eq:chi2}
\end{equation}
where $C_{ij}$ is the covariance matrix and
\begin{equation}
    \Delta \hat{P}_i(\bar{\tau}_{AP})
    =
    \hat{P}_{obs}(r_i)
    -
    \hat{P}_{th}(r_i,\bar{\tau}_{AP})
    \label{eq:deltaP}
\end{equation}
is the residual between the observed and the predicted pairwise kSZ. The pairwise kSZ covariance matrix is estimated with bootstrap resampling of tracer catalog. For each bootstrap realization, we randomly resample the catalog with replacement and recompute the pairwise kSZ signal. We generate 1,000 bootstrap realizations by resampling the galaxy catalog with each realization contains the same number of objects as the original catalog.

We also calculate the probability-to-exceed (PTE) as a goodness of fit statistic. It is defined as the probability of obtaining a $\chi^2$ value larger than the best fit value:
\begin{equation}
    {\rm PTE}
    =
    \int_{\chi^2_{min}}^{\infty}
    p(\chi^2)\,{d}\chi^2 ,
\end{equation}
where $p(\chi^2)$ is the chi-square probability distribution. The detection significance is quantified by the signal to noise ratio(SNR) of the best fit value,
\begin{equation}
    {\rm SNR}
    =
    \sqrt{
    \sum_{i,j}
    \hat{P}_{th}(\hat{\tau}_{AP})
    C^{-1}_{ij}
    \hat{P}_{th}(\hat{\tau}_{AP})
    }.
\end{equation}
Here $\hat{\tau}_{AP}$ is the best fit value of the mean effective optical depth.

In our analysis we use, and compare, two complementary approaches to infer the mean effective optical depth. In the first approach, we use the theoretical pairwise velocity predicted by linear theory, $V_{th}$ obtained from the {\it Planck} best fit, described in Eq.~(\ref{eq:Vhat_theo}). In the second, we use the machine learning reconstructed velocities, $v_{rec\parallel}$, to estimate the pairwise velocity statistic using Eq.~(\ref{eq:Vhat}). In both cases, the optical depth is obtained by fitting the corresponding velocity template to the observed pairwise kSZ signal. The AP-filtered effective optical depth is typically of order $10^{-4}$ for the groups and clusters traced by the massive LRG samples. It measures the average LOS electron density associated with the tracer population.

\subsection{Velocity-weighted density profile}
\label{sec:velocity_profile}
We  measure the mean stacked density profile for groups and clusters in the sample using the reconstructed
LOS peculiar velocities and the measured kSZ temperatures. Since LOS peculiar velocities can be positive or negative, an  unweighted stacking of kSZ temperatures would average the signal to zero. We therefore weight each AP-filtered CMB temperature by the reconstructed velocity and estimate the velocity-weighted kSZ
profile as
\begin{equation}
    T_{AP}^{kSZ}(\theta)
    =
    -\frac{1}{r}
    \frac{\sigma_{v,rec}}{c}
    \frac{
    \sum_i T_{AP,i}(\theta)
    \left(v_{rec \parallel,i}/c\right)
    }{
    \sum_i
    \left(v_{rec \parallel,i}/c\right)^2
    },
    \label{eq:velocity_weighted_stack}
\end{equation}
where $T_{AP,i}(\theta)$ is the AP-filtered temperature around the
$i^{th}$ galaxy with aperture size $\theta$, $\sigma_{v,rec}$ is the standard deviation of the reconstructed LOS velocities, and $r$ is the correlation coefficient between the reconstructed and true
velocities. The factor $\sigma_{v,rec}/r$ corrects for the normalization and imperfect correlation of the reconstructed velocity
field. For the velocity calibration factor, we adopt $r=0.91$, following the validation of the reconstructed velocities on simulations \cite{Gong:2026xld}. The adopted value of $r$ directly rescales the normalization of the velocity-weighted kSZ profile. Therefore, uncertainty in $r$ should be interpreted as a multiplicative calibration uncertainty on the profile amplitude. In future work, we will quantify the sensitivity of the stacked profile to the velocity calibration factor and propagate its uncertainty into the inferred profile amplitude. We apply this estimator in bins of aperture
radius from 1.6$^\prime$ to 5.6$^\prime$ with 0.5$^\prime$ bin distance to obtain the stacked kSZ profile. The uncertainty of the profile is estimated using spatial block resampling, which preserves correlations among galaxies within the same sky region. In the absence of velocity correlated foregrounds, this estimator provides a proxy for the AP-filtered projected optical depth profile of ionized gas around the LRG groups. 

\section{Results}
\label{sec:results}
In Sec.~\ref{sec:Phat_result}, we present the pairwise kSZ measurements using the estimator described in Sec.~\ref{sec:pairwise}. In Sec.~\ref{sec:Vhat_result}, we present the pairwise velocity measurements using the velocity reconstruction method described in Sec.~\ref{sec:velocity}. We then infer the mean effective optical depth using the procedure described in Sec.~\ref{sec:tau}, and finally present the velocity-weighted stacked kSZ profiles in Sec.~\ref{sec:stack_kSZ} using the estimator introduced in Sec.~\ref{sec:velocity_profile}.
\begin{figure*}
    \centering
    \includegraphics[width=\linewidth]{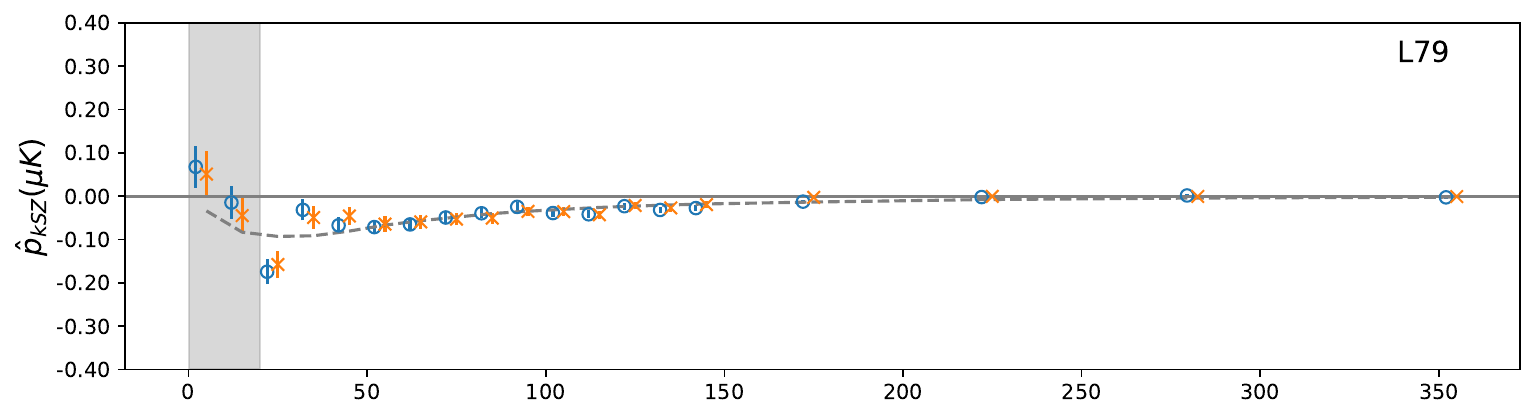}
    \includegraphics[width=\linewidth]{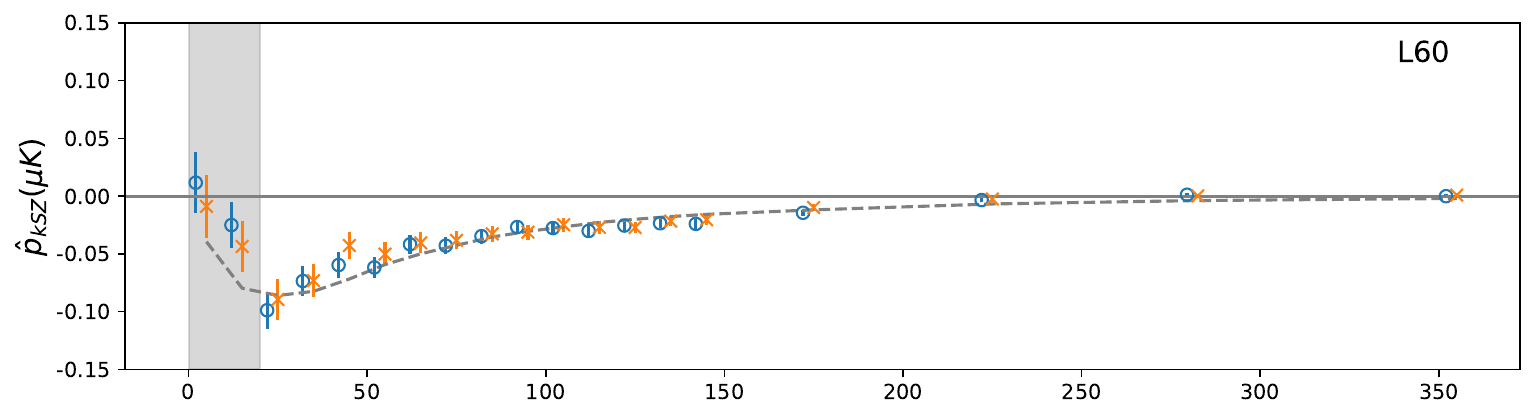}
    \includegraphics[width=\linewidth]{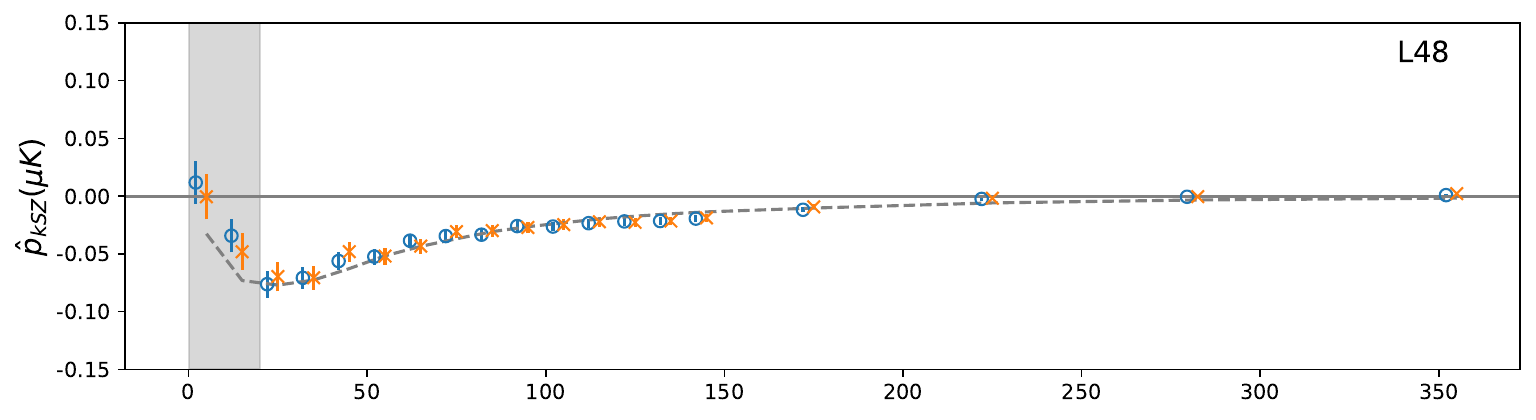}
    \includegraphics[width=\linewidth]{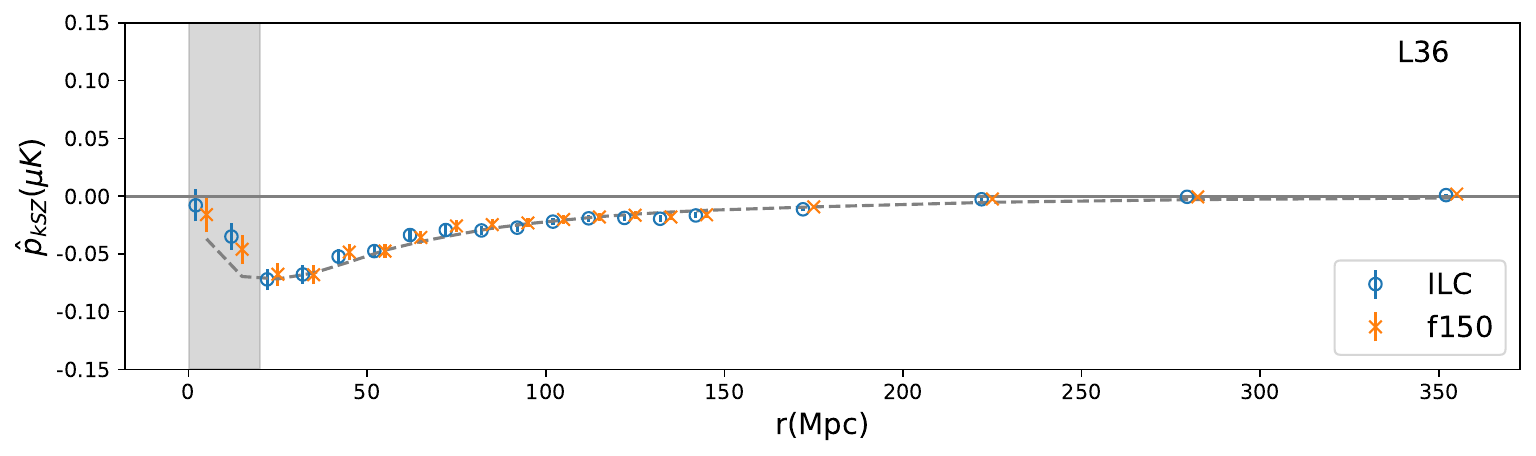}
    \caption{
    Pairwise kSZ measurements for the cumulative LRG luminosity bins (from bottom to top, the L36, L48, L60, and L79 samples) for the ACT DR7 ILC [blue circles] and f150 [orange crosses] maps. The gray shaded region indicates separations below 20Mpc, which are excluded from the optical depth fits. The dashed lines show the best-fit theoretical pairwise kSZ profiles for the ILC map obtained by fitting the effective optical depth using the theoretical pairwise velocity template. The 1$\sigma$ uncertainty estimated from a bootstrap analysis are also shown. \label{fig:phat_cumulative}}
\end{figure*}

\begin{figure*}
    \centering
    \includegraphics[width=\linewidth]{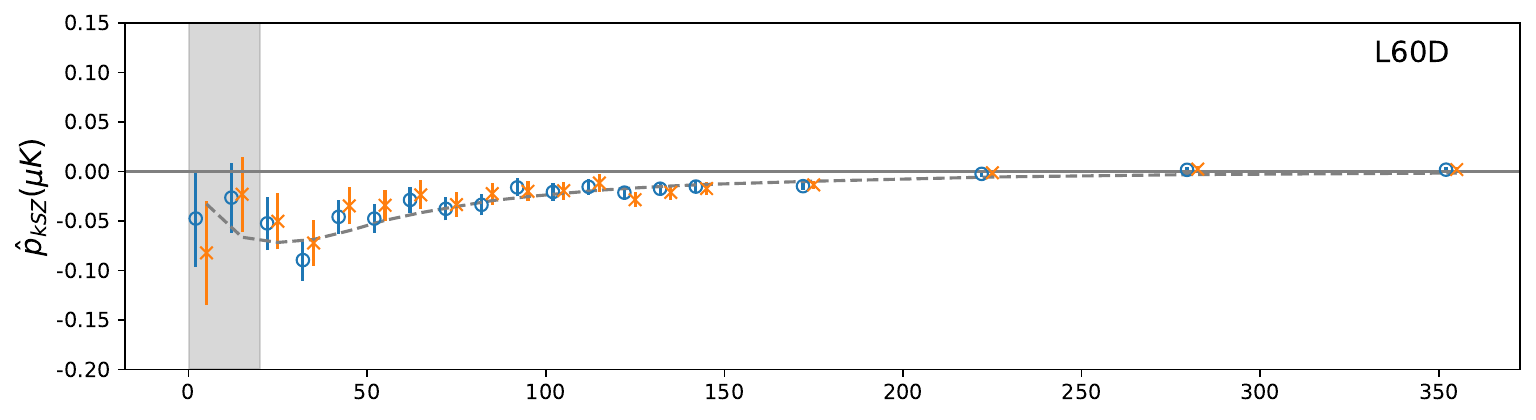}
    \includegraphics[width=\linewidth]{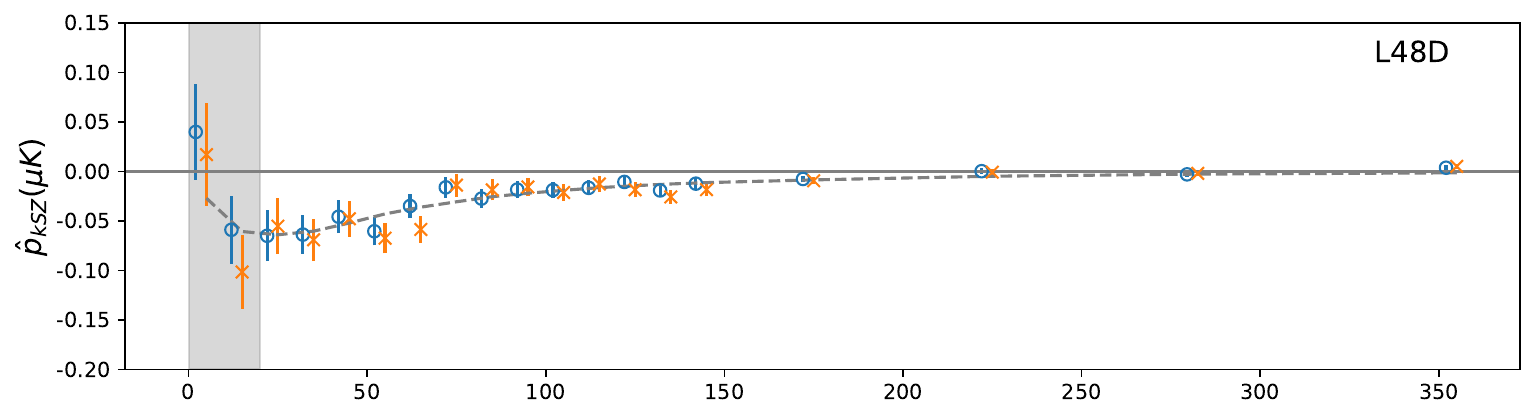}
    \includegraphics[width=\linewidth]{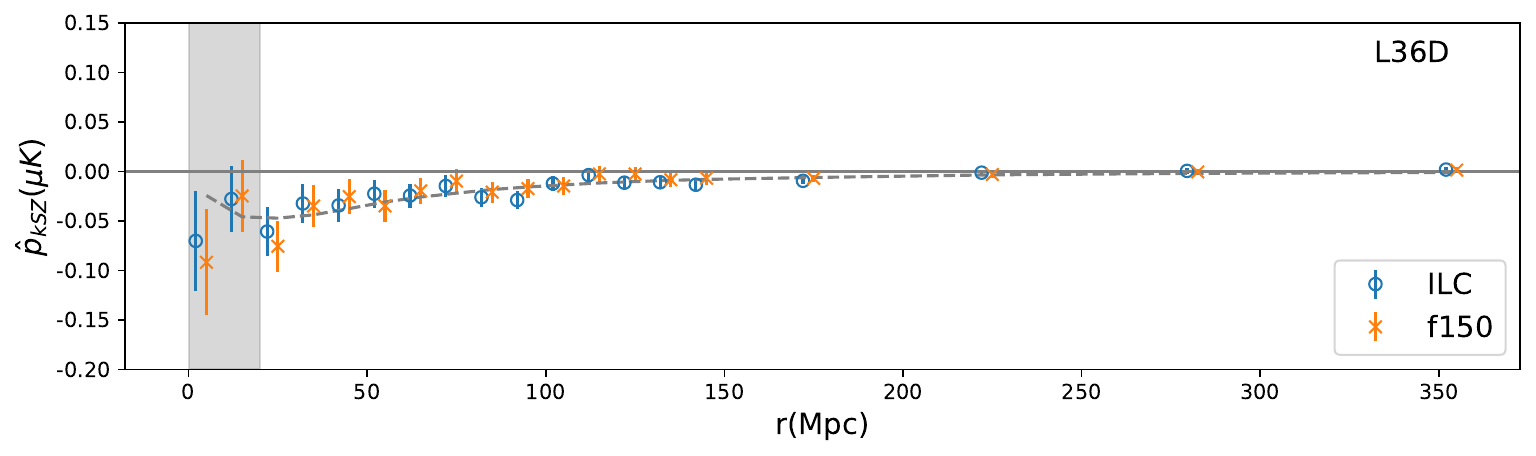}
    \caption{As in Fig.~\ref{fig:phat_cumulative} 
    but for the disjoint LRG luminosity bins: L36D [bottom], L48D [center] and L60D [top].}
    \label{fig:phat_discrete}
\end{figure*}

\subsection{Pairwise kSZ measurements}
\label{sec:Phat_result}

We first present the pairwise kSZ measurements derived from the ACT DR6 CMB maps and the DESI DR2 LRG samples. Fig.~\ref{fig:phat_cumulative} shows the results of four cumulative luminosity bins, L36, L48, L60, and L79. For each sample, we compare the ACT DR6 ILC map result with the f150 map result. We find that the two maps give consistent pairwise kSZ profiles with $1 \sigma$ uncertainty. The agreement between the ILC and f150 measurements in these cumulative bins provides a consistency test, indicating that the measurement is not driven by a strong frequency-dependent contamination. 

The shape of the signal is smoother for the lower luminosity bins, as the larger sample size gives smaller statistical uncertainties. All the four samples show a similar scale dependence where the pairwise signal reaches its peak amplitude around 25 Mpc and then gradually decays toward zero. We also find that the amplitude increases as luminosity increases. This is expected as the more luminous sample represents the sample with higher mass. 

In addition to the overall detection, the scale dependence of the measured profiles provides an important check on the physical origin of the signal. The pairwise kSZ signal is expected to be strongest on quasi-linear scales, where the infall velocities are significant, and to approach zero at large separations where the pairwise velocity vanishes. The consistency between maps, together with the smooth decay of the signal toward large scales, supports the interpretation that the measured temperature decrement is dominated by the kSZ signal rather than by residual thermal SZ, dust, or CIB contamination. A comparison with the 90GHz map would provide a further independent test of frequency-dependent contamination, which we leave to future work.

Figure~\ref{fig:phat_discrete} shows the corresponding measurements for the three disjoint luminosity bins, L36D, L48D, and L60D. These bins represent narrower luminosity ranges and therefore provide a useful check that the cumulative bin pairwise signal is not dominated by a single subset of galaxies. The disjoint samples again show a similar profiles as the cumulative ones. As expected, the uncertainties are larger than the cumulative samples because each disjoint bin contains fewer objects. The ILC and f150 results remain statistically consistent within $1 \sigma$ uncertainty, and the scale dependence is broadly compatible with the cumulative bin measurements. Since these discrete bins have similar sample size and the samples cover different luminosity and mass ranges, these measurements will also be useful for testing the dependence of the effective optical depth on halo mass.

\subsection{Pairwise velocity statistics from machine learning velocity reconstruction }
\label{sec:Vhat_result}

We estimate the LOS peculiar velocity for two  redshift-selected subsamples of the full LRG (L36) sample, $0.45<z<0.65$ and $0.65<z<0.95$, described in Sec.~\ref{sec:velocity} using the nonlinear velocity reconstruction method developed in \cite{Gong:2026xld}. 

\begin{figure*}
    \includegraphics[width=\columnwidth]{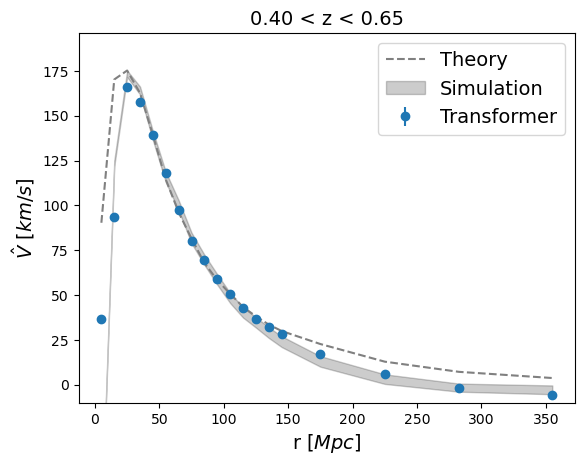}
    \includegraphics[width=\columnwidth]{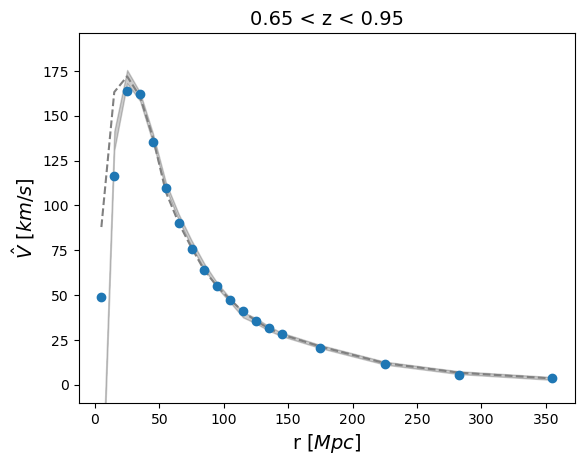}

    \caption{Pairwise velocity measurements in two redshift bins, $0.40<z<0.65$ [left] and $0.65<z<0.95$ [right] derived from the machine learning-reconstructed LOS peculiar velocities with the $1\sigma$ uncertainty estimated from bootstrap resampling. The shaded regions show the corresponding $1\sigma$ uncertainties for the AbacusSummit simulation. The {\it Planck} best fit theoretical pairwise velocity prediction [dashed gray line]  using Eq.~(\ref{eq:Vhat_theo}) are also included for comparison.}
    \label{fig:vhat_redshift}
\end{figure*}

The resulting pairwise velocity statistics are shown in Fig.~\ref{fig:vhat_redshift}. We also present the predictions from the AbacusSummit simulations. The simulation curve is computed from one representative realization of the simulation, while the shaded region shows the corresponding $1\sigma$ uncertainty estimated from the covariance of the 25 independent realizations provided in the simulation. In each realization, we use the same number of tracers as in the corresponding DESI L36 redshift bin, so that the simulation uncertainty includes a comparable contribution from sample size fluctuations. We randomly downsample the tracer catalog in each AbacusSummit realization to match the number of tracers in the corresponding DESI sample. The smaller uncertainty in the higher-redshift bin is due to its sample size being twice that of the lower-redshift bin.

In both redshift bins, the measured pairwise velocity is consistent with the profile shape of the pairwise kSZ momentum, corresponding to the expected coherent infall of galaxy pairs. The amplitude decreases toward larger separations and approaches zero, as expected when the velocity correlation becomes weak for distant pairs. 

In addition to serving as an input for the velocity-weighted density profile measurement, the reconstructed pairwise velocity statistic provides an important independent validation of the velocity reconstruction itself. The kSZ signal is proportional to the LOS momentum of free electrons, and therefore any interpretation of the stacked kSZ profile relies on the accuracy of the reconstructed peculiar velocities. By measuring the pairwise velocity directly from the reconstructed LOS velocities and comparing it with the expectation from simulations, we test whether the reconstruction recovers not only the one-point velocity distribution, but also the two-point coherent infall pattern that is relevant for large-scale structure.

The measured profiles are consistent with the simulation predictions on scales above approximately $45\,{\rm Mpc}$. Comparing the reconstructed pairwise velocity measurements with the AbacusSummit prediction over the 17 separation bins for $0.65<z<0.95$, we obtain $\chi^2=12$ for 17 degrees of freedom, indicating statistical consistency between the two. This agreement is particularly meaningful because the pairwise velocity is a direct tracer of gravitational growth. On large scales, galaxy pairs are expected to move coherently toward each other due to the growth of overdense regions. The amplitude and scale dependence of this infall signal are determined by matter clustering, tracer bias, and the growth rate of structure. Therefore, recovering the expected pairwise velocity profile demonstrates that the reconstructed velocity field retains the physical information needed to connect the observed galaxy distribution to the underlying cosmic velocity field. Although the theoretical and reconstructed pairwise velocity profiles show some differences at the largest separations, these scales contribute relatively little statistical weight to the optical depth fit discussed in the following section. 

The agreement at large separations provides the cleanest validation of the recovered velocity field. These scales are dominated by coherent large-scale flows, where linear theory and simulation predictions are expected to be more reliable. In contrast, smaller scales can be more strongly affected by nonlinear velocities, halo-scale motions, satellite dynamics, and reconstruction smoothing. For this reason, we interpret the consistency above $45\,{\rm Mpc}$ as the most relevant validation that the reconstructed velocities capture the cosmological velocity field.

The redshift split also provides a useful consistency test. The two redshift bins probe different cosmic epochs and are matched to the redshift ranges over which the velocity reconstruction models were trained and validated. The agreement between the reconstructed pairwise velocity and the simulation prediction in both bins indicates that the reconstruction remains stable across the redshift range used in this analysis. This is important because a redshift-dependent bias in the velocity reconstruction could mimic or obscure redshift evolution in the kSZ signal. The consistency between the two bins therefore supports the interpretation that the subsequent differences, or lack of differences, in the stacked kSZ profiles are not primarily driven by the velocity reconstruction. 

\begin{table*}
\centering
\setlength{\tabcolsep}{6pt}
\renewcommand{\arraystretch}{1.25}
\begin{tabular}{|c|c|c|c|c|c|c|c|c|}
\hline
\multirow{2}{*}{\begin{tabular}[c]{@{}c@{}}Tracer\\ sample\end{tabular}} & 
\multicolumn{4}{c|}{$f150$} & 
\multicolumn{4}{c|}{ILC} \\
\cline{2-9}
 & $\bar{\tau}_{AP}$ ($\times 10^{-4}$) & $\chi^2_{\text{min}}$ & PTE & SNR & 
   $\bar{\tau}_{AP}$ ($\times 10^{-4}$) & $\chi^2_{\text{min}}$ & PTE & SNR \\
\hline

L36 & 
0.44 $\pm$ 0.03 & 18 & 0.33 & 13.3 & 
0.47 $\pm$ 0.03 & 21 & 0.19 & 15.1 \\

L48 & 
0.47 $\pm$ 0.04 & 22 & 0.17 & 11.6 & 
0.49 $\pm$ 0.04 & 17 & 0.39 & 13.5 \\

L60 & 
0.48 $\pm$ 0.05 & 16 & 0.49 & 9.6 & 
0.53 $\pm$ 0.05 & 23 & 0.11 & 12.0 \\

L79 & 
0.51 $\pm$ 0.07 & 24 & 0.09 & 7.2 & 
0.55 $\pm$ 0.07 & 39 & 0.01 & 8.5 \\

\hline

L36D & 
0.30 $\pm$ 0.08 & 8 & 0.96 & 4.0 & 
0.31 $\pm$ 0.07 & 12 & 0.74 & 4.5 \\

L48D & 
0.45 $\pm$ 0.07 & 26 & 0.05 & 6.4 & 
0.41 $\pm$ 0.06 & 19 & 0.30 & 6.4 \\

L60D & 
0.40 $\pm$ 0.07 & 14 & 0.55 & 5.4 & 
0.44 $\pm$ 0.07 & 14 & 0.58 & 6.8 \\

\hline

\end{tabular}
\caption{
Summary of the mean AP-filtered effective optical depth estimates,  $\bar{\tau}_{AP}$, derived from the pairwise kSZ, using the [left] 150GHz and [right] ILC ACT maps, through comparison with theoretical linear pairwise velocity correlations predicted for the cluster mass ranges given for each LRG tracer sample in Table~\ref{tab1} and the {\it Planck} best fit cosmological parameters,  using separations $r >$ 20Mpc. The minimum $\chi^2$ (for 17 degrees of freedom), probability-to-exceed (PTE) the observed value and signal-to-noise (SNR) for each theoretical fit with the best fit optical depth are also given.}
\label{tab:tau_results}
\end{table*}

\begin{table*}
\centering
\setlength{\tabcolsep}{6pt}
\renewcommand{\arraystretch}{1.25}
\begin{tabular}{|c|c|c|c|c|c|c|c|c|}
\hline
\multirow{2}{*}{\begin{tabular}[c]{@{}c@{}}Tracer\\ sample\end{tabular}} & 
\multicolumn{4}{c|}{$f150$} & 
\multicolumn{4}{c|}{ILC} \\
\cline{2-9}
 & $\bar{\tau}_{AP,rec}$ ($\times 10^{-4}$) & $\chi^2_{\text{min}}$ & PTE & SNR & 
   $\bar{\tau}_{AP,rec}$ ($\times 10^{-4}$) & $\chi^2_{\text{min}}$ & PTE & SNR \\
\hline
L36z1 & 
0.46 $\pm$ 0.08 & 15 & 0.56 & 6.1 & 
0.52 $\pm$ 0.07 & 19 & 0.26 & 7.9\\
L36z2 & 
0.49 $\pm$ 0.05 & 13 & 0.65 & 9.7 & 
0.46 $\pm$ 0.05  & 15 & 0.51 & 10.0\\ \hline
L36$_{sub}$ & 
0.48 $\pm$ 0.04 & 15 & 0.52 & 12.5 & 
0.49 $\pm$ 0.04 & 16 & 0.43 & 14.2 \\
L48$_{sub}$ & 
0.50 $\pm$ 0.05 & 18 & 0.35 & 11.0& 
0.51 $\pm$ 0.04 & 17 & 0.41 & 12.2\\
L60$_{sub}$ & 
0.52 $\pm$ 0.06 & 13 & 0.69 & 8.9 & 
0.57 $\pm$ 0.06 & 18 & 0.32 & 10.4\\
L79$_{sub}$ & 
0.56 $\pm$ 0.10 & 19 & 0.21 & 5.6 & 
0.61 $\pm$ 0.10 & 26 & 0.04 & 6.5\\ \hline
L36D$_{sub}$ & 
0.37 $\pm$ 0.08 & 11 & 0.84 & 4.7 & 
0.37 $\pm$ 0.08 & 10 & 0.89 & 5.1\\
L48D$_{sub}$ & 
0.51 $\pm$ 0.09 & 23 & 0.12 & 6.1& 
0.40 $\pm$ 0.08 & 13 & 0.51 & 4.8\\
L60D$_{sub}$ & 
0.44 $\pm$ 0.09 & 13 & 0.69 & 5.0& 
0.48 $\pm$ 0.08 & 15 & 0.52 & 6.0\\
\hline

\end{tabular}
\caption{As in Table~\ref{tab:tau_results} but presenting the mean AP-filtered effective optical depth estimates, $\bar{\tau}_{AP,rec}$, obtained by comparison of the kSZ pairwise with the nonlinear pairwise velocity correlations from machine learning-based velocity reconstruction. Results are shown for two $z$-selected subsamples of L36, $0.4<z<0.65$ (L36z1), $0.65<z<0.95$ (L36z2), and the corresponding subsamples (sub) of all the samples in Table~\ref{tab:tau_results} with redshift restricted to $0.4<z<0.95$.}
\label{tab:tau_results2}
\end{table*}

\subsection{$\tau$ estimates using pairwise velocity}
\label{sec:tau}
In this section, we infer the mean effective optical depth using two complementary approaches. First, we fit the pairwise kSZ measurements presented in Section~\ref{sec:Phat_result} using the theoretical pairwise velocity template from Eq.~\ref{eq:Vhat_theo}. Second, we also combine the same pairwise kSZ measurements with the machine learning reconstructed pairwise velocities presented in Section~ \ref{sec:Vhat_result} to infer the mean effective optical depth.

For both the cumulative and the disjoint bins, we use the pairwise profiles to constrain the mean effective optical depth by fitting the measured pairwise kSZ measurement to the theoretical pairwise velocity template described in Sec.~\ref{sec:pairwise}, assuming the {\it Planck} bestfit cosmology, using separations $r > 20 Mpc$. 

The estimated optical depth values inferred here should be interpreted as AP-filtered effective optical depths rather than the total integrated optical depths. Table \ref{tab:tau_results} summarizes the best fit values and detection significances.

The ILC and f150 likelihoods are generally consistent with each other within $1 \sigma$ uncertainty, supporting the interpretation that the inferred optical depth is robust to the choice of CMB temperature map. We find that, for most of the luminosity bins, the ILC results are consistently larger than the f150 results. This is expected as a result of the slightly different beam sizes of the two maps.

 For the cumulative samples, both the f150 and ILC maps give significant detections. The SNR of the ILC measurements are higher than the corresponding f150 measurements, but the two sets of constraints are broadly consistent within the statistical uncertainties. The optical depth values of the cumulative bins also increase with luminosity. This trend is qualitatively consistent with the expectation that more luminous samples occupy more massive halos and therefore have larger electron densities.

The SNR decreases toward higher luminosity bins. This is expected because the higher luminosity bins contain fewer tracers and therefore have larger statistical uncertainties. The L36 and L48 samples provide the highest significance, with SNR values larger than 10 in both the f150 and ILC maps. The highest significance detection is obtained for the cumulative L36 sample using the ILC map, with SNR=15.1. This improves on the strongest detection reported in the previous ACT DR6 x DESI DR1 analysis \cite{Gong:2025ffw}, where SNR = 9.3 for the same L36 luminosity cut and the ILC map. The higher SNR is driven primarily by the larger DESI DR2 LRG sample size and the resulting reduction in statistical uncertainty. The f150 measurement for L36 also gives a high significance detection with SNR = 13.3, indicating that the improved significance is not specific to the ILC map alone.

For the disjoint luminosity bins, we also obtain significant detections of the pairwise kSZ signal, although with lower SNR than the cumulative samples because of the smaller sample sizes. The ILC measurements give SNR = 4.5, 6.4, and 6.8 for L36D, L48D, and L60D, respectively, while the corresponding f150 measurements are consistent within the statistical uncertainties. The inferred AP derived optical depths are also broadly consistent with an increasing trend with luminosity or halo mass, with the lowest-luminosity disjoint bin giving the smallest optical depth. Therefore, the disjoint bin measurements provide a useful consistency check on the cumulative bin results and suggest a luminosity or halo mass dependence of the effective optical depth.

We also jointly analyze the pairwise kSZ measurements with the pairwise velocity estimates from the machine learning-based velocity reconstruction for the equivalent samples to obtain purely empirically reconstructed velocity-derived optical depth estimates, $\bar{\tau}_{AP,rec}$. We study both the two redshift selected subsamples of L36: $0.4<z<0.65$ (L36z1) containing  482,087 LRGs with $\langle z\rangle =$ 0.55, and $0.65<z<0.95$ (L36z2) containing  971,889 LRGs with $\langle z\rangle =$ 0.80, as well as the subsamples of all the seven luminosity selected samples with redshift restricted to $0.4<z<0.95$. Indicative of the subsamples, $L36_{sub}$ contains $\sim$76\% of the full L36 sample (which spans $0<z<1.5$), but has a similar average redshift, $\langle z \rangle$=0.74 (vs 0.76 for L36).

To obtain the likelihood for $\bar{\tau}_{AP,rec}$ we use an analogous approach to that in (\ref{eq:chi2}) but with $\hat{P}_{rec}(r) = \bar{\tau}_{AP,rec}\hat{V}(r)$ replacing $\hat{P}_{th}(r)$. While the covariance matrix is now dependent on both the kSZ  and peculiar velocity measured covariances, we find that for the ACT and DESI data, the uncertainties in the CMB measurements are dominate over those from the peculiar velocities and it is reasonable to approximate the total covariance as the kSZ covariance to obtain the $\bar{\tau}_{AP,rec}$ likelihood.

The results are summarized in Table~\ref{tab:tau_results2}. For both maps, the optical depths derived from the reconstructed velocity are consistent between the two redshift bins, L36z1 and L36z2, within $1 \sigma$ 
uncertainty;  we find no evidence of redshift evolution for the AP-filtered effective optical depth within the current statistical uncertainties. The results of the two subsamples are also consistent with that of the overall result for the two samples combined, L36$_{sub}$. 

\begin{figure*}
    \centering
    \includegraphics[width=\linewidth]{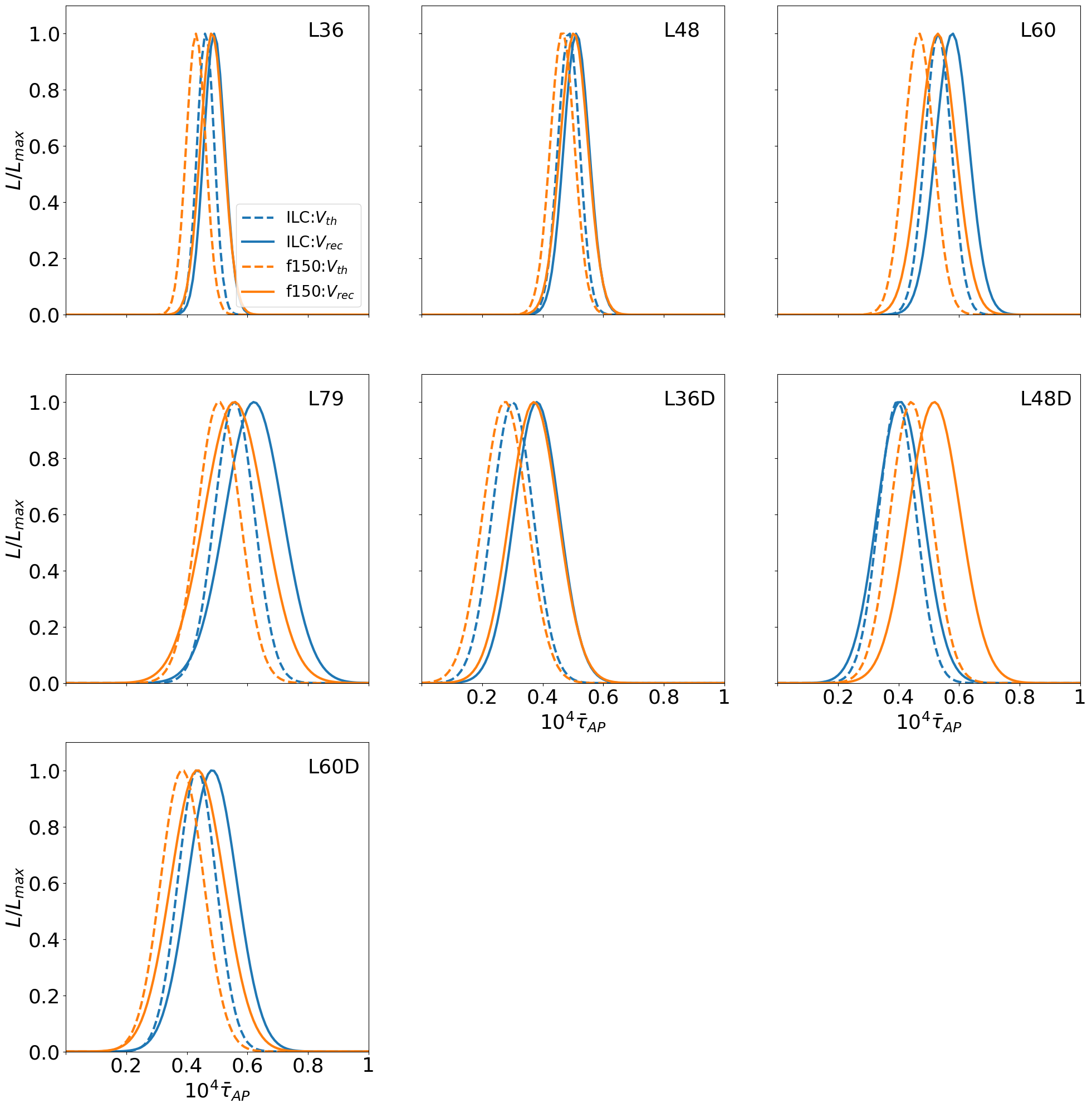}
    \caption{Normalized likelihoods for the AP-filtered mean effective optical depth inferred from the pairwise kSZ measurements from the the ACT DR6 ILC [blue] and f150 [orange] maps. The $\tau_{AP,rec}$ likelihoods using the velocity pairwise correlation from reconstruction [full lines] are shown alongside $\bar{\tau}_{AP}$ from fitting to the theoretical prediction using the {\it Planck} best fit cosmology [dashed lines]. The likelihoods are normalized assuming a uniform prior over the plotted domain. }
    \label{fig:tau_likelihood}
\end{figure*}

Figure~\ref{fig:tau_likelihood} compares the likelihoods for the cumulative and disjoint bins for both the  optical depths inferred from the pairwise velocity obtained from the theoretical prediction using {\it Planck} best fit cosmology and the DESI LRG velocity reconstruction for the $0.4<z<0.95$ subsample. 

For all subsamples of the luminosity-selected samples restricted to the redshift range $0.4<z<0.95$, the f150 and ILC constraints also agree well with each other within the $1\sigma$ uncertainties. This agreement is important because the ILC and f150 maps have different foreground properties and map-making details, and therefore provides a useful check that the inferred optical depths are not driven by a map-specific systematic. 

The reconstructed velocity-derived optical depths are also consistent with those inferred using the Planck best-fit pairwise velocity template. The agreement indicates that restricting the redshift range and replacing the theoretical pairwise velocity prediction with the pairwise velocity measured directly from the DESI reconstructed velocity field do not significantly shift the inferred mean optical depth. When the combined L36 sample is used, the empirical reconstructed-velocity constraints reach SNR = 12.5 for f150 and SNR = 14.2 for ILC, slightly lower than the constraints using the theoretical {\it Planck} velocity-template analysis, as expected given the smaller sample size. 

These results demonstrate how the combination of the measured pairwise kSZ signal and the machine learning reconstructed pairwise velocity provides a fully empirical way to estimate the mean effective AP-filtered optical depth. Although this method is not fully model independent,  it does not require assuming the theoretical pairwise velocity amplitude from a fiducial cosmology. Instead, the velocity information is extracted directly from the observed DESI LRG density field, based on the machine learning model trained on synthetic data. This provides a complementary and more empirical estimate of the optical depth. The consistency between the optical depth estimates  supports both the robustness of the kSZ measurement and the validity of the reconstructed velocity field for measuring the baryonic gas around massive LRG halos.

\begin{figure}
\includegraphics[width=\columnwidth]{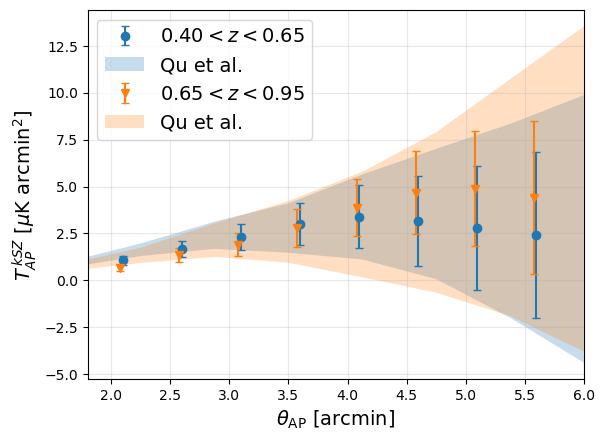}
    \caption{
    Velocity-weighted stacked cluster density profiles measured from the ACT DR6 ILC map around DESI DR2 LRGs in two redshift bins. The [blue circles] show the result for $0.40<z<0.65$ (L36z1), and the [orange triangles] show the result for $0.65<z<0.95$ (L36z2). Error bars show the $1\sigma$ uncertainty estimated from spatial block covariance. We also show the corresponding DESI LRG measurements from \cite{Qu:2026zyh}, which analyzed a similar redshift range using a linear velocity reconstruction method, for comparison.}
    \label{fig:stack_ksz}
\end{figure}

\subsection{Stacked kSZ density profiles}
\label{sec:stack_kSZ}

The velocity-weighted stacked kSZ profile provides spatial information about the ionized gas distribution around LRG groups, complementary to the mean effective optical depth constraints from the pairwise kSZ measurement. By weighting the AP-filtered temperature by the reconstructed LOS velocity, the kSZ signal adds coherently while velocity independent foregrounds largely cancel, allowing us to probe how far the gas extends beyond the central halo region.

We finally measure the velocity-weighted stacked cluster density profile defined in Eq.~\ref{eq:velocity_weighted_stack} for the DESI DR2 LRGs using the temperature measured from the ACT DR6 ILC map. A similar analysis using linearly reconstructed velocities was presented in \cite{Qu:2026zyh}. This measurement provides a complementary view of the gas distribution to the pairwise kSZ statistic discussed above. While the pairwise estimator uses temperature differences between galaxy pairs to isolate the coherent infall, the stacked profile uses the reconstructed LOS velocity of each galaxy to combine the CMB temperature field at the galaxy positions. Because the sign of the kSZ temperature depends on the sign of the peculiar velocity, weighting by the reconstructed velocity prevents the signal from averaging to zero. In the absence of contaminants that correlate with the reconstructed velocity field, this estimator is proportional to the optical depth profile of the ionized gas around the LRG sample. 

Figure~\ref{fig:stack_ksz} shows the resulting kSZ profiles in the two redshift bins, $0.40<z<0.65$ and $0.65<z<0.95$ of the L36 sample. At small aperture radii, the measured signal is lower because the filter encloses only the central part of the projected gas distribution and is more sensitive to beam smoothing and AP attenuation effects \citep{Gong:2023hse}. 
The signal then increases with aperture size and peaks around $\theta_{AP}\simeq 4$--$5$ arcmin, indicating that a substantial fraction of the kSZ weighted gas signal is distributed over several arcminutes around the LRGs. This suggests that the ionized gas around massive galaxies is spatially extended rather than being confined to the inner most halo region.

This radial behavior is consistent with the broader picture emerging from recent DESI$\times$ACT kSZ measurements, where stacked kSZ profiles are interpreted as direct probes of the projected electron distribution around galaxies. In particular, the flattening of the AP filtered profile at larger aperture radii suggests that the signal is not dominated only by gas in the central galaxy or the innermost halo, but receives a substantial contribution from gas distributed over circumgalactic and group scale distances. This interpretation was first highlighted for DESI LRGs in \cite{RiedGuachalla:2025byu} and is subsequently supported by a joint kSZ and CMB lensing analysis that constrained the gas distribution and gas fraction \cite{Hadzhiyska:2025mvt}. More recent measurements using DESI DR2 LRGs \cite{Qu:2026zyh} and BGS galaxies \cite{Hadzhiyska:2026uyq} also recover extended stacked kSZ profiles across different tracer populations. This agreement supports the interpretation that the ionized gas associated with massive galaxies is spatially extended, as feedback and gravitational heating redistribute baryons beyond the central halo region. The profiles of the two redshift bins are also statistically consistent within 1$\sigma$ uncertainties. Within the current uncertainties, we find no evidence for a strong evolution of the AP-filtered kSZ profile across the redshift range probed by the L36 sample. 

The profile measured here is expected to be robust to contamination from the thermal SZ effect and the cosmic infrared background. These signals do not change sign with the LOS peculiar velocity, whereas the kSZ temperature does. Therefore, contaminations that are uncorrelated with the reconstructed velocity will cancel when averaged with positive and negative velocities. We test this by repeating the measurement after randomly shuffling the reconstructed velocities among the catalog and by stacking at random sky locations. Both tests are consistent with a null signal within the statistical uncertainties, indicating that the measured profile is not generated by correlations in the temperature map.

We also compare our measurements with the recent DESI LRG stacked kSZ profile results from \cite{Qu:2026zyh}, which used the linear reconstruction method as discussed in Sec.\ref{sec:velocity}. While both analyses use DESI DR2 LRGs, the LRG samples are selected in different ways and are therefore not identical. We use different sample selections, including additional luminosity and redshift cuts. Since the two LRG samples are nevertheless very similar, we compare our measurements with the \cite{Qu:2026zyh} results in a similar redshift range. We find that the measured profiles are broadly consistent within the statistical uncertainties. This agreement provides a useful external cross check of the velocity-weighted stacking measurement. In particular, it suggests that the inferred extended kSZ profile is not driven by a specific choice of velocity reconstruction method or sample definition, but is a robust feature of massive DESI LRG halos. Small differences between the profiles can naturally arise from the different LRG selections, redshift distributions, and velocity reconstruction. Given these analysis differences and the current uncertainties, we do not interpret the comparison as evidence for a significant discrepancy. Instead, the consistency supports the interpretation that massive DESI LRG halos host extended ionized gas whose kSZ signal is detected over several arcminutes. We obtain SNRs of 11 and 10 for the L36z1 and L36z2 samples, respectively, compared with 10 and 9 in \cite{Qu:2026zyh} for samples spanning similar redshift ranges. The detection significances are therefore comparable, despite our sample sizes are approximately 20\% smaller due to the luminosity cut. This suggests that the improved velocity reconstruction may partially compensate for the reduced sample size, although a direct comparison is limited by differences in sample selection. For the velocity calibration factor, we adopt $r$=0.91, following its validation on simulations \cite{Gong:2026xld}. This factor directly affects the normalization of the stacked kSZ profile, while the resulting SNR also depends on the tracer sample, velocity weighting, CMB noise, and covariance estimates. A same-sample comparison between the linear and machine learning reconstructions, together with a sample matched recalibration of $r$, will be an important future validation for better understanding the difference in SNR. In addition, the relative difference between the two profiles is consistent with the trend found in \cite{Gong:2026xld}, where the impact of using different velocity reconstruction methods was explicitly tested. In that work, profiles obtained with linear velocity reconstruction were found to be higher than those obtained with the machine learning based velocity reconstruction for scales below 3 arcmins. This is qualitatively consistent with the comparison shown here. Therefore, the observed difference is not unexpected and can be naturally explained by the different velocity reconstruction methods, in addition to the slightly different LRG sample selections and analysis choices.

\section{Conclusion}
\label{sec:conclusion}
In this work, we present measurements of the kSZ signal and peculiar velocity fields for groups and clusters centered on LRGs using DESI DR2 cataogs  and ACT DR6 CMB temperature maps. Specifically we obtain measurements of the pairwise kSZ momentum, pairwise velocity correlation, and velocity-weighted stacked kSZ profiles. The combination of the high resolution ACT DR6 maps with the DESI DR2 spectroscopic LRG sample, which is over twice the size of the DR1 sample studied in previous work \cite{Gong:2023hse}, enables high significance detections of the kSZ signal from massive galaxy halos over a broad redshift range.

The pairwise kSZ signal is measured for four cumulative luminosity bins and three disjoint luminosity bins. The measurements from the ACT DR6 ILC and f150 maps are broadly consistent with each other, providing an important check that the signal is not dominated by strongly frequency-dependent foreground contamination. We find that the inferred AP-filtered effective optical depth increases with luminosity, consistent with the expectation that more luminous LRG samples occupy more massive halos and contain more ionized gas.  

We also inferred the pairwise velocity statistic using machine learning reconstruction of the LRG LOS peculiar velocities. The reconstructed pairwise velocity correlations in two redshift bins show the expected coherent infall signal and agree with the prediction from simulations and the theoretical prediction based on the {\it Planck} best fit cosmology. 

By fitting the measured pairwise kSZ signal to the theoretical pairwise velocity template using the {\it Planck} best fit cosmology, we obtained constraints on the AP filtered mean effective optical depth, $\bar{\tau}_{\rm AP}$. 
The best detection is obtained for the L36 sample using the ACT DR6 ILC map, with SNR=15.1. 
This represents an improvement over the previous ACT DR6 $\times$ DESI DR1 pairwise kSZ measurement, mainly due to the larger DESI DR2 LRG sample and the corresponding reduction in statistical uncertainty. The consistency between the ILC and f150 results further supports the robustness of the optical depth constraints.

By combining the measured pairwise kSZ signal with the reconstructed pairwise velocity, we obtain an empirical estimate of the AP-filtered effective optical depth. For the L36 subsamples covering the same redshift ranges as the simulated lightcone datasets used to train the machine learning model, the optical depths derived from the reconstructed velocity are consistent between the two redshift bins and are also with the value inferred for the combined sample. This suggests that there is no evidence for strong redshift evolution in the AP-filtered effective optical depth within the current uncertainties. We also infer the mean effective optical depth for all the luminosity selected sample but restricted the redshift to $0.4 < z <0.95$. The optical depth measurements are also consistent with those obtained for the full sample using the {\it Planck} best fit model. The linear reconstruction is an intermediate step of the machine learning framework and could be used independently to repeat the analysis. A systematic same-sample comparison between the linear and machine learning reconstructions is left to future work.

Finally, we measured velocity-weighted stacked kSZ profiles in two redshift bins. The amplitude increases with aperture radius and become relatively flat at larger apertures, suggesting that the kSZ signal receives contributions from ionized gas distributed over extended circumgalactic regions rather than being confined only to the central galaxy. The two redshift bins are statistically consistent within the current uncertainties, and we find no evidence for strong redshift evolution of the AP filtered profile across the range probed by the L36 sample. This stacked measurement provides a complementary view of the gas distribution to the pairwise kSZ statistic. The pairwise measurement constrains an effective optical depth while the stacked profile probes the scale dependence of the projected electron distribution.

Together, these measurements demonstrate the power of combining DESI spectroscopic galaxies with ACT CMB maps to study both baryonic gas and cosmic velocity fields. The pairwise kSZ detection, the reconstructed pairwise velocity measurement, and the velocity-weighted stacked kSZ profiles provide mutually consistent evidence for extended ionized gas around massive LRG halos and for coherent large scale motions traced by these galaxies. Applying the same framework to additional LSS tracers from DESI, as well as other upcoming surveys including from Rubin LSST and the Euclid and Roman Space Telescopes, and future CMB data sets, such as Simons Observatory and CCAT, will enable more detailed tests of the baryon distribution, feedback physics, and the growth of structure.

\begin{acknowledgments}

The work of YG and RB is supported by NSF grant AST-2206088, NASA grant 22-ROMAN11-0011 and NASA grant 12-EUCLID12-0004. EMV and JEM are supported by the Department of Energy grant DE-SC0010007.


This research used resources of the National Energy Research Scientific Computing Center (NERSC), a U.S. Department of Energy Office of Science User Facility located at Lawrence Berkeley National Laboratory, operated under Contract No. DE-AC02-05CH11231 using NERSC award HEP-ERCAPmp107

This material is based upon work supported by the U.S. Department of Energy (DOE), Office of Science, Office of High-Energy Physics, under Contract No. DE–AC02–05CH11231, and by the National Energy Research Scientific Computing Center, a DOE Office of Science User Facility under the same contract. Additional support for DESI was provided by the U.S. National Science Foundation (NSF), Division of Astronomical Sciences under Contract No. AST-0950945 to the NSF’s National Optical-Infrared Astronomy Research Laboratory; the Science and Technology Facilities Council of the United Kingdom; the Gordon and Betty Moore Foundation; the Heising-Simons Foundation; the French Alternative Energies and Atomic Energy Commission (CEA); the National Council of Humanities, Science and Technology of Mexico (CONAHCYT); the Ministry of Science, Innovation and Universities of Spain (MICIU/AEI/10.13039/501100011033), and by the DESI Member Institutions: \url{https://www.desi.lbl.gov/collaborating-institutions}. Any opinions, findings, and conclusions or recommendations expressed in this material are those of the author(s) and do not necessarily reflect the views of the U. S. National Science Foundation, the U. S. Department of Energy, or any of the listed funding agencies. The authors are honored to be permitted to conduct scientific research on I'oligam Du'ag (Kitt Peak), a mountain with particular significance to the Tohono O’odham Nation.

\end{acknowledgments}

\section*{DATA AVAILABILITY}
The data corresponding to the figures in this paper are available at https://doi.org/10.5281/zenodo.20790055.

\clearpage
\bibliography{draft}

\end{document}